\UseRawInputEncoding
\documentclass[aps,hidelinks,twocolumn, prl, superscriptaddress,float fix,preprintnumbers]{revtex4-2}
\usepackage{graphicx}
\usepackage{dcolumn}
\usepackage{orcidlink}
\usepackage{caption}
\usepackage{float}
\usepackage{subcaption}
\usepackage{csquotes}
\usepackage{float}
\usepackage{amsfonts}
\usepackage{amsmath}
\usepackage{amsmath,amssymb}
\usepackage[italicdiff]{physics}
\usepackage{bm}
\usepackage{mathtools}
\usepackage{enumitem}
\usepackage{tensor}
\usepackage{wrapfig}
\usepackage{footnote}
\usepackage{here, comment}
\usepackage{bm}
\usepackage{booktabs}

\begin{document}

\preprint{RUP-25-11}
\preprint{YITP-25-66}

\title{Quantum correlation beyond entanglement:\\ Holographic discord and multipartite generalizations}

\author{Takato Mori\,\orcidlink{0000-0001-6442-875X}}
\email{takato.mori@yukawa.kyoto-u.ac.jp}
\affiliation{Department of Physics, Rikkyo University, 
3-34-1 Nishi-Ikebukuro, Toshima-ku, Tokyo 171-8501, Japan}
\affiliation{Perimeter Institute for Theoretical Physics, Waterloo, Ontario N2L 2Y5, Canada}
\affiliation{Center for Gravitational Physics and Quantum Information, Yukawa Institute for Theoretical Physics, Kyoto University, Kitashirakawa Oiwakecho, Sakyo-ku, Kyoto 606-8502, Japan}

\date{\today}

\begin{abstract}

While entanglement is a cornerstone of quantum theory and holography, quantum correlations arising from superposition, such as quantum discord, offer a broader perspective that has remained largely unexplored in holography. 
We construct gravity duals of quantum discord and classical correlation. In both holographic systems and Haar random states, discord exceeds entanglement, revealing an additional quantum correlation linked to the Markov gap and non-distillable entanglement, suggesting holographic states are intrinsically non-bipartite. In black hole setups, discord can increase despite decoherence and persists beyond the sudden death of distillable entanglement. Motivated by the holographic formula, we define reflected discord---an optimization-free boundary quantity based on reflected entropy---which remains effective even outside the holographic regime. We also propose several multipartite generalizations of correlation measures. It includes holography-inspired correlations based on multi-entropy, which are shown to be UV-finite and reduce to bipartite measures in the bipartite limit. These results provide new tools for quantifying quantum correlations beyond entanglement in strongly coupled many-body systems and offer a novel approach to multipartite correlation measures.

\end{abstract}

\maketitle

\section{Introduction}
Quantum correlations that originate from quantum superposition are a defining feature of quantum theories. 
While entanglement is the most celebrated example both 
in quantum information~\cite{Horodecki:2009zz,Chitambar:2018rnj,Ma:2023ecg} and quantum gravity~\cite{Ryu:2006bv,Maldacena:2013xja,Maldacena:2001kr, VanRaamsdonk:2009ar, Czech:2012be, Lewkowycz:2013nqa, Pastawski:2015qua, Jacobson:2015hqa, Witten:2018zxz},
mixed-state correlations extend beyond it.
For bipartite pure states, quantum correlation is equivalent to entanglement, and it is uniquely quantified by entanglement entropy~\cite{Donald_2002}.
In contrast, quantum correlations of mixed states are more diverse. Even within the plathora of entanglement measures, there are multiple distinct measures for bipartite mixed states such as entanglement of formation~\cite{Hill:1997pfa,Wootters:1997id}, distillable entanglement~\cite{Bennett:1995tk,Bennett:1995ra,Bennett:1996gf}, and squashed entanglement~\cite{Christandl_2004}.
In holographic dualities, thanks to the entanglement wedge reconstruction, a mixed state has a counterpart in the bulk~\cite{Cotler:2017erl,Czech:2012bh}. Corresponding to various entanglement measures proposed in quantum information, their gravity duals have been proposed occasionally by relying on a geometric optimization~\cite{Mori:2024gwe,Takayanagi:2017knl,Dutta:2019gen,Cheng:2019aqf,Tamaoka:2018ned,Wen:2021qgx}. Still it remains elusive both in quantum information and quantum gravity how two subsystems are correlated in mixed states.

While entanglement is a genuinely quantum phenomenon, mixed states have an even richer structure. 
Even separable mixed states can exhibit non-classical correlations.
Unlike classical systems where we have a designated choice of basis, we can superpose the basis in quantum settings, which leads to a correlated state due to quantum coherence. For example,
\begin{equation}
	\rho_{AB} = \frac{1}{2} \qty(\dyad{00}+\dyad{1+})\label{eq:discordant}
\end{equation}
has zero entanglement but nonzero non-classical/quantum correlation.
Compared to a classically correlated state like $\frac{1}{2}(\dyad{00}+\dyad{11})$, the basis of the second subsystem is partly superposed as $\ket{+}=\frac{1}{\sqrt{2}}(\ket{0}+\ket{1})$. 

Quantum discord is a measure that captures this type of quantum correlation without entanglement~\cite{Henderson_2001,PhysRevLett.88.017901,Zurek_2000}. 
It is defined as a difference between the quantum mutual information and the classical counterpart by conditioning via an optimal subsystem measurement that maximizes the classical information gain. 
This quantity quantifies quantum correlation including entanglement. Indeed, for a pure state, quantum correlation is entanglement so the discord equals entanglement entropy. 
The discord becomes particularly interesting for mixed states, where some states have non-classical correlation other than entanglement as seen in \eqref{eq:discordant}.
It is recently gaining attention in high/low energy physics and cosmology as well as quantum information. See~\cite{Han:2024ugl,Afik:2022dgh,Guo:2020rwj,Martin:2021znx,Sugiyama:2024equ,Matsumura:2020uyg} for recent developments.
Despite its great interest, computing discord is challenging because one must optimize over all measurements on one subsystem. As a result, analytical computation has been done only for
small systems or Gaussian states~\cite{Galve_2011,Zhu_2018,Giorda_2010,Adesso_2010,Ali_2010,ma2015quantum,He_2015}.
Numerical computation is also challenging due to infinitely many choices of measurements~\cite{Wu_2009,LANG_2011,Huang_2014}.

In addition, while quantum discord is considered to capture a wider class of quantum correlations compared to entanglement, it is not apparent if it is quantitatively larger than or equal to some mixed-state entanglement measure. In fact, it is very subtle as a related quantity called the geometric discord can be smaller than an entanglement measure~\cite{Rana_2012} and it is not straightforward to provide an operational definition of quantum discord as it can be increased by a local operation~\cite{PhysRevA.86.034101,Torun:2024djb}.

To explore a further correlation structure of holographic spacetime beyond entanglement and overcome the computational difficulty in quantum discord, we consider quantum discord in holographic settings. Based on an earlier study by the author~\cite{Mori:2024gwe} on the holographic locally accessible information, we find the gravity dual of the quantum discord. Using the formula, we are able to quantify quantum correlation beyond entanglement, which was missing in conventional studies of holographic entanglement measures.

Unlike previous studies on the quantum discord focusing on finite-dimensional (few-body) systems or Gaussian states, this holographic approach circumvents the computational difficulty by employing geometric optimization. Holographic states are expected to be defined in a strongly interacting field theory, which is something that has barely been explored in the context of quantum discord due to its large dimensions and interactions. 

Our proposed formula also provides additional quantitative evidence that quantum discord probes quantum correlation more than entanglement. We find that, for both holographic and Haar random mixed states, the quantum discord is strictly greater than the squashed entanglement, which is a faithful entanglement measure, while the quantum discord always coexists with entanglement. This implies that there always exists non-entanglement quantum correlation (NEQC) in holography and random states. Furthermore, we find the NEQC is related to the Markov gap between one of the subsystems of interest and the purifying subsystem. Since the Markov gap was proposed as a diagnosing quantity for the tripartite entanglement~\cite{Hayden:2021gno,Zou:2020bly}, our analysis indicates that the tripartite entanglement implies non-entanglement quantum correlation and vice versa. Based on the author's previous work~\cite{Mori:2024gwe}, we also find that the NEQC equals the non-distillable entanglement. This further suggests its operational interpretation.

Besides exploring the notion of quantum correlation beyond entanglement, a generalization of the bipartite correlation to a multipartite setting has also garnered much interest recently. While the notion of the multipartite entanglement is not unique and there are various distinct proposals~\cite{Horodecki:2024bgc, Ma:2023ecg, Walter:2016lgl, Choi:2022lge, Gadde:2024jfi, Memmen:2023ftr, Dur:2000zz, Gadde:2022cqi, Gadde:2023zzj, Penington:2022dhr, Gadde:2023zni, Gadde:2024taa, Harper:2024ker, Iizuka:2025ioc, Iizuka:2025caq, Yuan:2024yfg, Basak:2024uwc, Balasubramanian:2024ysu, Ju:2024hba, Akers:2019gcv}, we argue that the classical correlation and quantum discord have natural generalizations to the multipartite settings. We discuss five different possibilities including a generalization from a holographic perspective using the so-called multi-entropy~\cite{Gadde:2023zzj,Penington:2022dhr}.

This paper is organized as follows. In Section~\ref{sec:prelim}, we review the classical correlation and the quantum discord, which are the main quantities discussed in this paper. 
In Section~\ref{sec:proposal}, we present our main proposal for the bulk duals of classical correlation and quantum discord and examine their properties. 
In Section~\ref{sec:ex}, we compute the holographic correlation measures in various setups in AdS$_3$/CFT$_2$ and Haar random states. 
In Section~\ref{sec:NEQC}, we define the non-entanglement quantum correlation (NEQC) and present its gravity dual. Its relation with the Markov gap and distillable entanglement is clarified. In Section~\ref{sec:reflected}, the boundary dual of the holographic correlation measures is 
proposed based on the reflected entropy. 
In Section~\ref{sec:multi}, we propose five different multipartite generalizations of these correlation measures including a holographic one based on the multi-entropy. In Section~\ref{sec:concl}, we conclude with a summary and future outlook in quantum information and gravity. Table~\ref{tab:corr} provides a list of notations for the correlation measures examined in this paper.
Appendix~\ref{app:BTZ} presents a holographic computation 
for the thermofield double state, whereas Appendix~\ref{app:hol-BH} deals with the Gibbs state. Appendix~\ref{app:rank2} presents the boundary dual computation 
for rank-two two-qubit states.

\section{Preliminaries}\label{sec:prelim}
\subsection{Classical and quantum correlation}
Among various measures proposed to evaluate classical and quantum correlations, we start by defining classical correlation, also known as the locally accessible information, and quantum correlation called quantum discord. For further reviews of classical and quantum correlations, see~\cite{Adesso:2016ygq,Modi_2012,datta2008studies,Bera_2017,C_LERI_2011}.

Conventionally, the correlation between two states is quantified by mutual information. The total mutual correlation between $A$ and $B$ in  $\rho_{AB}$ is defined through the mutual information, which is given by
\begin{equation}
	I(A:B)=S_A+S_B-S_{AB}=S_A-S(A|B),
\end{equation}
where $S_{A,B}=-\Tr\rho_{A,B}\log\rho_{A,B}$ is the entanglement entropy and $S(A|B)=S_{AB}-S_B$ is the quantum conditional entropy~\footnote{We occasionally denote $S_A$ by $S(A)$ in this paper.}.  Throughout the paper, all entropies are natural-log based.
Focusing on the subsystem $A$, the mutual information is defined as the entropy of $A$ minus the entropy conditioned on $B$ by subtracting $S(B)$. Instead, one can condition the entropy by measuring $B$. This leads to the definition of \emph{classical correlation}, which is another type of mutual information:
\begin{equation}
	J(A|B) = \max\qty[S_A - \sum_i p_i S(\rho_i^A)],
	\label{eq:cc}
\end{equation}
where $\rho_i^A$ is the reduced density matrix on $A$ after a POVM with outcome $i$ on $B$ and $p_i$ denotes the corresponding probability. The maximization is taken over all possible POVM measurements supported only on $B$. 
$J(A|B)$ can be written as a convex combination of relative entropies, and thus non-negative~\footnote{See~\cite{Mori:2024gwe} for instance.}.

In~\cite{Henderson_2001,Vedral_2003}, $J(A|B)$ is identified as classical correlation as the conditioning by the measurement transforms the marginal quantum state to a probability distribution, which is classical. Indeed, when the measured subsystem $B$ is classical, that is, the state is a quantum-classical (QC) state
\begin{equation}
	\rho_{QC}= \sum_i p_i \rho_i^A\otimes \dyad{i}_B,
    \label{eq:qc-state}
\end{equation}
two definitions of mutual information coincide. In general, we have $I\ge J$, agreeing with the intuition that the total correlation includes the classical correlation as part.

The \emph{quantum discord} is defined as the difference between the total correlation and the classical correlation~\cite{Henderson_2001,PhysRevLett.88.017901,Zurek_2000}:
\begin{equation}
	D(A|B)=I(A:B)-J(A|B)=S_B-S_{AB}+\min \sum_i p_i S(\rho_i^A).
    \label{eq:q-discord}
\end{equation}
From the monotonicity of the relative entropy, this quantity is non-negative and the non-negativity of $J(A|B)$  implies $D(A|B)\le I(A:B)$.
As expected, the quantum discord reduces to entanglement entropy for a bipartite pure state and vanishes for QC states. However, it is not necessarily equal to entanglement; for example, a discordant state like~\eqref{eq:discordant} has nonvanishing quantum discord while its entanglement is zero.

While these quantities are useful to probe classical correlation and quantum correlation broader than entanglement, they involve optimization in their definition, which makes their analytical and numerical computation hard. On a different note, there are diverse past works indicating interesting connections to mixed-state quantum computation, state merging, entanglement distribution, quantum cryptography, quantum thermodynamics, and quantum metrology~\cite{Datta_2008,Madhok_2011,Cavalcanti_2011,Streltsov_2017,MADHOK_2012,Pirandola_2014,PhysRevA.81.062103,Girolami_2015}. Nevertheless, these operational interpretations of quantum discord has been debated as they are often context-dependent~\cite{Yu_2013,Jose:2024zgb,Haque:2025xjj}. After all, it is still an ongoing debate if the discord can be a necessary and sufficient resource in some resource theory~\cite{Daki__2012,Yadin_2016,Jebarathinam:2024xex}. 
In the later sections, we argue that holographic discord avoids the difficulty in the optimization and may have operational interpretation in relation to the entanglement distillation and the Markov gap.

\subsection{Holographic entanglement measures}
In this subsection, we review several holographic quantities relevant to several entanglement quantities.
We focus on time-reflection symmetric cases in this paper although a generalization can be done similarly as entanglement entropy or entanglement wedge cross section.
Furthermore, we only consider the leading-order contribution in Newton's constant $G_N$. Any subleading corrections $o(1/G_N)$ are neglected.

In AdS/CFT, any boundary quantity has a bulk realization. Entanglement entropy is not an exception.
Entanglement entropy between a boundary subregion $A$ and its complement is holographically given by the area of the codimension-two minimal surface homologous to $A$, divided by $4G_N$~\cite{Ryu:2006bv}. A natural extension of holographic entanglement entropy to mixed states is given by the entanglement wedge cross section (EWCS) in the bulk~\cite{Takayanagi:2017knl}:
\begin{equation}
	E_W(A:B)=\min_{\Gamma_{A:B}} \frac{\mathrm{Area}(\Gamma_{A:B})}{4G_N},
\end{equation}
where the minimization is over all possible codimension-two surface $\Gamma_{A:B}$ that divides the entanglement wedge of $AB$ into two regions, each homologous to $A$ or $B$. The EWCS is proposed to be dual to multiple distinct boundary quantities, including the entanglement of formation~\cite{Mori:2024gwe}, entanglement of purification~\cite{Takayanagi:2017knl,Caputa:2018xuf}, the balanced partial entanglement~\cite{Wen:2021qgx}, and odd entropy~\cite{Tamaoka:2018ned}.

Another key quantity is the squashed entanglement~\cite{Christandl_2004}
\begin{equation}
	E_{sq}(A:B)\equiv \frac{1}{2}\inf I(A:B|C), 
\end{equation}
where the infimum is taken over all extensions such that $\Tr_C\rho_{ABC}=\rho_{AB}$.
This quantity is known to be a faithful measure of entanglement, which means the measure is zero if and only if the state is separable, and satisfies various desirable properties, making itself one of the most promising bipartite entanglement measures.

In holography, it was proposed according to a geometric optimization or the linearity of entanglement entropy that its infimum is achieved by a pure extension with empty $C$~\cite{Cheng:2019aqf,Umemoto:2018jpc,Takayanagi:2017knl}. Namely,
\begin{equation}
    E_{sq}(A:B)=\frac{1}{2}I(A:B)\qq{in holography.}
    \label{eq:hol-sq}
\end{equation}
We will often compare various correlation measures to this to see their discrepancy with entanglement.

\section{Holographic classical and quantum correlations}\label{sec:proposal}
In this section, we propose the gravity dual of classical and quantum correlations in terms of holographic entanglement entropy and EWCS and examine its properties.

\subsection{Gravity duals}
Let us consider the gravity dual of the classical correlation $J(A|B)$~\eqref{eq:cc}. We first note that the classical correlation is equivalent to the locally accessible information by definition. Thus, all we need is the gravity dual of the locally accessible information.

It is conjectured in~\cite{Mori:2024gwe} that the locally accessible information (LAI) $J(A|B)$ is dual to
\begin{equation}
	J_W(A|B)=S_A-E_W(A:C),
	\label{eq:cc-hol}
\end{equation}
which we call the \emph{holographic classical correlation} in this paper.
The subsystem $C$ is the complement or the purifier of $AB$. 
Note that from the definition of the EWCS, $J_W(A|B)$ does not depend on the choice of $C$.
This holographic formula is equivalent to the duality between the entanglement of formation $E_F$ and the EWCS through the Koashi-Winter relation, which says
\begin{equation}
    J(A|B)=S_A-E_F(A:C).
\end{equation}
In general, identifying the optimal measurement is a hard task. We resolve this issue and derive \eqref{eq:cc-hol} under three assumptions:
\begin{enumerate}[label=(\alph*)]
  \item The path integral saddle for the average measurement outcome is dominated by a single bulk geometry and the entropy of the post-measurement state is outcome-independent at leading order.
  \item A boundary POVM measurement on a subsystem only backreacts within its entanglement wedge.
  \item A disentangled basis localized along the minimal surface exists.
\end{enumerate}

The first assumption is supported by the approximate orthogonality of macroscopically distinct holographic states through the gravitational path integral~\footnote{We stress that backreacted saddles discussed in holographic measurements in earlier literature such as~\cite{Antonini:2022sfm,Numasawa:2016emc} and saddles for $J(A|B)$ and $D(A|B)$ in our paper are defined differently. In the former cases, the focus is a post-measurement state. As in the standard AdS/CFT correspondence, its gravity dual is given by some saddles. In contrast, in our case, the saddle is the one dominating the average over the measurement outcomes. Since in both $J(A|B)$ and $D(A|B)$, the post-measurement entropy is weighted by the measurement probability $\{p_i\}_i$, contributions from the post-measurement states which is macroscopically distinct from the original state are typically suppressed even if each post-measurement state may admit a saddle-point description.}. Then, via the surface/state correspondence, we introduce a complete basis spanning the $e^{S_B}$-dimensional Hilbert space of $B$ along $\gamma_B$. It is given by a random disentangled basis along the minimal surface in a tensor network picture. The measurement probability is expected to be flat over the exponentially many outcomes, corresponding to the same saddle-point geometry but with a different `microstate' labled by the local basis along $\gamma_B$.
See~\cite{Mori:2024gwe} for the illustration.

The second assumption stems from the entanglement wedge reconstruction. While there might be a backreaction in the complementary region in the bulk, the smoothness of the bulk due to the Einstein equation forces a backreaction in the entanglement wedge of $A$. Such a contribution is likely to be subdominant in $J(A|B)$ as the weight $p_i$ is exponentially suppressed as discussed in the first assumption~\footnote{Here we implicitly assume that the number of such backreacted saddles with the same residual entropy is not exponentially large. An exception is a saddle with an end-of-the-world brane-like object localized on the boundary of the entanglement wedge.}. 

Another possible caveat to the second assumption is the insertion of a charged operator. For example, suppose one can perform an operation breaking the boundary global symmetry. Starting with a state with charge $q$ in a subregion $B\subseteq A^c$, where $A^c$ is a complement of another subregion $A$, an insertion of charge $(-q)$ in $A$ leads to a Wilson line connecting two charges across the entanglement wedge of $A$. When such a global constraint is present, the entanglement wedge reconstruction is more subtle. However, this is unlikely a problem as causality prohibits such a spooky action at a distance~\cite{Mori:2024gwe}. Furthermore, we are interested in the leading-order regime so these charges are either irrelevant for $O(1/G_N)$ (i.e. matter charges) or backreact and change the structure of the entanglement wedge itself according to the quantum extremal surface prescription. 

The third assumption is trivially satisfied in the random tensor network model~\cite{Mori:2024gwe} although in more realistic setups it is less trivial~\cite{Hayden:2020vyo,Wei:2022cpj,Hung:2024gma}.

Since the locally accessible information is the same as the classical correlation, \eqref{eq:cc-hol} gives its holographic formula. 
It is non-negative as $E_W$ is upper bounded by the entanglement entropy.

The \emph{holographic quantum discord} directly follows from the holographic classical correlation. It is given as
\begin{equation}
	D_W(A|B)=I(A:B)-J_W(A|B)=S_B-S_{AB}+E_W(A:C),
\end{equation}
where $C$ is an arbitrary purification partner of $AB$.

In the light of the AdS/CFT duality, which relates the bulk to a strongly-coupled conformal field theory, this formula predicts the amount of quantum discord in the regimes where previous studies cannot reach. Until now, the study of quantum discord is limited to a few-qubit systems or Gaussian cases whereas our formula is expected to be valid in the strongly-coupled quantum field theories.

\subsection{Properties}
Let us now review the properties of classical and quantum correlations. We will see shortly that all the properties are satisfied by the gravity duals as well.

\begin{figure}
	\centering
	\includegraphics[width=0.23\textwidth]{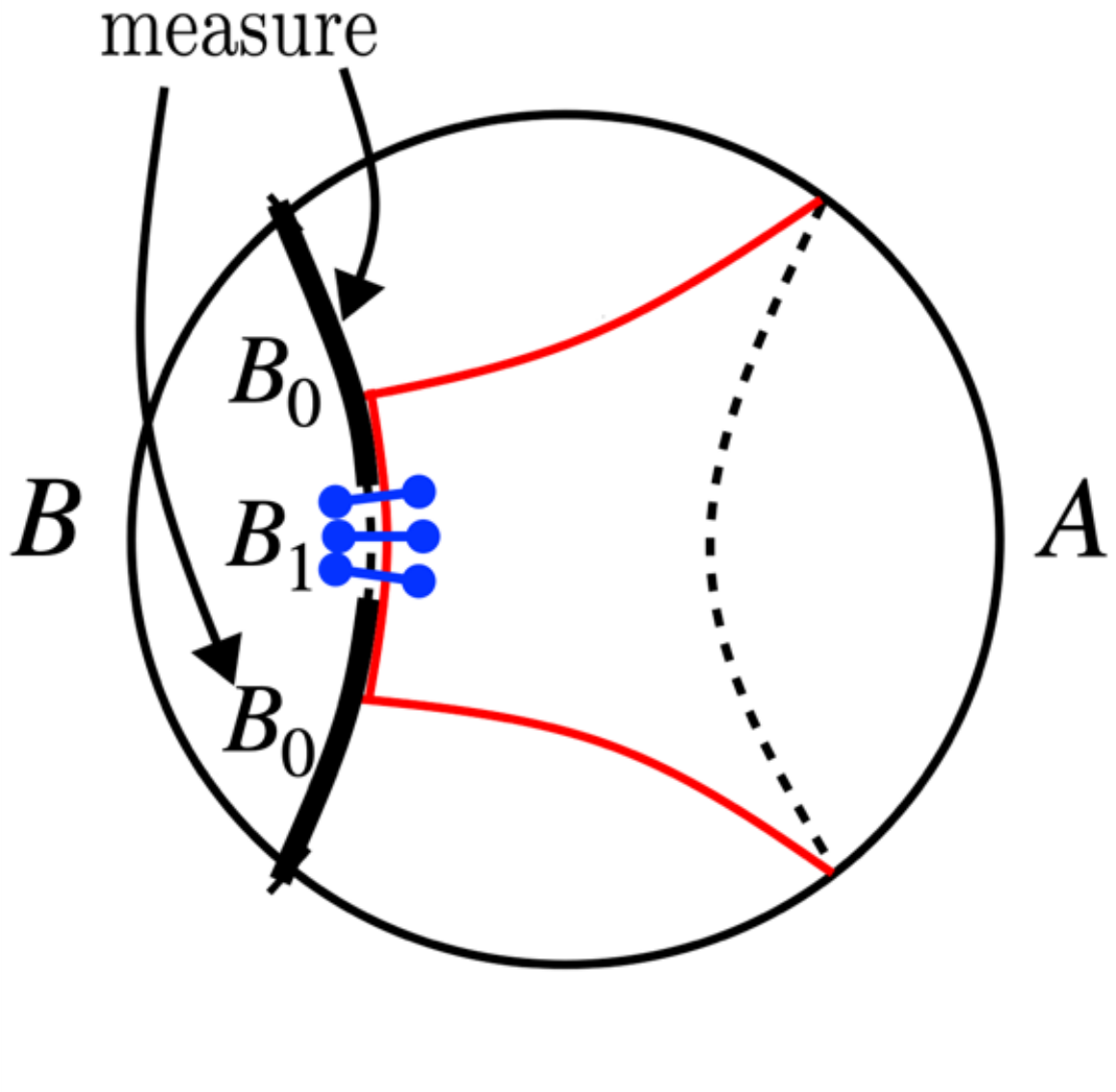}
	\caption{By measuring $B_0$ and send its outcome to $A$, the entanglement wedge of $A$ becomes touching to the entanglement wedge of $B$. Along the overlapping surface $B_1$, one can distill $\mathrm{Area}(B_1)/(4G_N)$ EPR pairs (denoted as blue lines). 
    }
	\label{fig:JwED}
\end{figure}

\paragraph{Classical correlation.---}
The classical correlation satisfies the following properties~\cite{Vedral_2003,Henderson_2001,Streltsov_2011}:
\begin{enumerate}
	\item $J(A|B)\le I(A:B)$,
	\item $\rho=\rho_A\otimes\rho_B \Rightarrow J(A|B)=0$,
	\item $J(A|B)=S_A=S_B$ for pure $\rho_{AB}$.
	\item $J(A|B)\le \min(S_A,S_B)$,
	\item $J(A|B)$ is invariant under local unitary transformations,
	\item $J(A|B)$ is non-increasing under local operations.
\end{enumerate}
All of these properties are satisfied by the proposed gravity dual $J_W(A|B)$. 
Property (1) is equivalent to the non-negativity of the holographic discord, which follows from
\begin{align}
	D_W(A|B)&=S_{AC}-S_C+E_W(A:C) \nonumber\\
	&\ge S_{AC}-S_C+(S_A+S_C-S_{AC})/2 \nonumber\\
	&=(S_A+S_{AC}-S_C)/2 \nonumber\\
	&=\frac{I(A:B)}{2}\ge 0, \label{eq:DwEsq}
\end{align}
where $ABC$ forms a pure state.
The first inequality follows from the fact that the EWCS is lower bounded by half of mutual information~\cite{Freedman:2016zud,Takayanagi:2017knl}.
Property (2) immediately follows from property (1). 
Property (3) follows from $E_W(A:C)=0$ for pure states. 
When $S_A<S_B$, property (4) is evident from the non-negativity of the EWCS. However, when $S_B<S_A$, it is less trivial.
For the proof, it is useful to recall the alternative operational interpretation of $J_W(A|B)$. In~\cite{Mori:2024gwe}, $J_W(A|B)$ is identified to be equal to (or at least lower bound) the one-way distillable entanglement between $A$ and $B$, namely, the number of maximally entangled (EPR) pairs that can be distilled along the minimal surface of $B$, denoted by $\gamma_B$, by measuring a portion of $B$ (Fig.~\ref{fig:JwED}). Property (4) follows since the number of EPR pairs distillable in this protocol is upper bounded by $S_B$.
Property (5) follows from the definition of entropy and boundary duals of EWCS. To see property (6), recall that any local operation can be implemented as a unitary acting on the system and ancillae and tracing over the ancillae. Since adding the ancilla enlarges the entanglement wedge, it suffices for the proof to show the monotonicity under the partial trace. 
$J(A|B^\prime)<J(A|B)$ with $B^\prime \subseteq B$ is equivalent to $E_W(A:C^\prime)>E_W(A:C)$ with $C^\prime\supseteq C$, which follows from the entanglement wedge nesting~\cite{Akers:2016ugt,Czech:2012bh,Wall:2012uf}. The monotonicity with respect to $A$ is more nontrivial.

\begin{figure}
	\centering
	\includegraphics[width=0.35\textwidth]{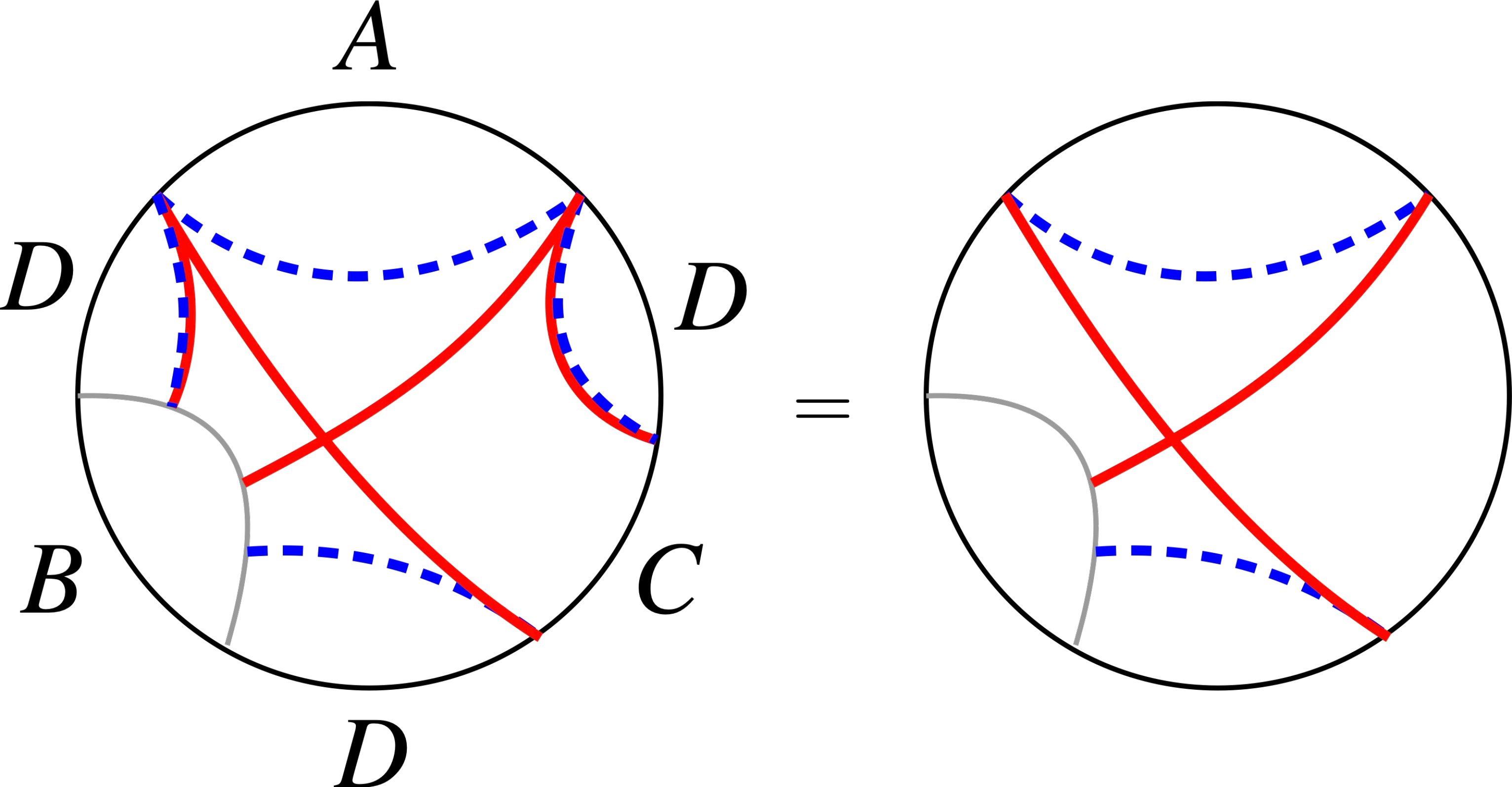}
	\caption{
    Geometric illustration that $F(A,C,D)\equiv S_{AC}-E_W(AC:D)-S_A+E_W(A:CD)\ge 0$. The solid red segments (counted positively) and the dashed blue segments (counted negatively) together form a closed loop whose total area is non-negative by extremality of holographic entanglement entropy and EWCS.
    }
	\label{fig:mono-geom}
\end{figure}

The monotonicity $J(AC|B)\ge J(A|B)$ is equivalent to
\begin{equation}
	F(A,C,D)\equiv S_{AC}-E_W(AC:D)- S_A+E_W(A:CD)\ge 0,
    \label{eq:mono-J}
\end{equation}
where $ABCD$ forms a pure state. This can be proven geometrically and the proof is presented~\cite{Louisia:2025bxz}. A nontrivial example of the geometrical proof is illustrated in Fig.~\ref{fig:mono-geom}. This monotonicity relation is a novel inequality involving the EWCS. Equivalently, this relation implies that the difference of either the EWCS or the Markov gap $h(A:B)$, which is dual to $2E_W(A:B)-I(A:B)$ in holography~\cite{Hayden:2021gno,Zou:2020bly}, lower bounds the conditional entropy or conditional mutual information $I(B:C|D)\equiv S_{BD}+S_{CD}-S_{BCD}-S_D$, respectively: 
\begin{align}
	E_W(AC:D)-E_W(A:CD)&\le S(C|A)\\
	h(AC:D)-h(A:CD)&\le I(B:C|D).
\end{align}

\paragraph{Quantum discord.---}
Let us now examine the properties of quantum discord~\cite{Henderson_2001,PhysRevLett.88.017901,Zurek_2000}.
\begin{enumerate}
	\item $0\le D(A|B)\le I(A:B)$
	\item $D(A|B)\le S_B$
	\item $D(A|B)=0$ if and only if $\rho_{AB}$ quantum-classical
	\item For pure states, $D(A|B)=S_A=S_B$
	\item $D(A|B)$ is invariant under local unitary transformations
	\item $D(A|B)$ is monotone under local operation on $A$
\end{enumerate}
Holographic discord also satisfies these properties as expected. Property (1) follows from $0\le J_W\le I$ and property (2) follows from $E_W(A:C)\le S_C=S_{AB}$.
To see property (3), we first argue that geometrically separable states are geometrically product states. By ``geometrically," we mean neglecting any subleading order in $G_N$. 
In holography one conjectures $E_F(A:B)=E_W(A:B)$~\cite{Mori:2024gwe} and $E_{sq}(A:B)=I(A:B)/2$~\cite{Cheng:2019aqf,Umemoto:2018jpc,Takayanagi:2017knl}. Since these vanish if and only if $A$ and $B$ are geometrically disconnected, any separable state must correspond to a disconnected wedge in the bulk, which implies holographic separable states are product states (to the leading order).
Since a QC state $\rho_{AB}=\sum_i \rho_i^A\otimes \dyad{i}_B$ is a separable state, it must be a product state in holography if it exists. Hence, what we need to show is $D_W(A|B)=0$ if and only if $I(A:B)=0$ in holography. $(\Rightarrow)$: From $E_W\ge I/2$, we have $D_W(A|B)\ge I(A:B)/2$. This means $D_W(A|B)=0$ implies $I(A:B)=0$. $(\Leftarrow)$: $I(A:B)=0$ implies $E_W(A:C)=S_A$. Then, $D_W(A|B)=I(A:B)=0$. This concludes the holographic proof of property (3).
Property (4) follows from $E_W(A:C)=S_{AB}=0$ for pure $AB$. Property (5) follows from the same reason as the invariance of $J(A|B)$. To show property (6), we need to show $D_W(AC|B)\ge D_W(A|B)$. 
In fact, this is equivalent to the monotonicity of the classical correlation \eqref{eq:mono-J}. Given its proof, property (6) is shown.

It is worth noting that from property (3), one can easily see quantum discord vanishes for any two parties of the GHZ state $\ket{\mathrm{GHZ}}=\frac{1}{\sqrt{d}}\sum_{i=1}^d \ket{iii}_{ABC}$ while it does not for the (generalized $n$-qubit) W state $\ket{\mathrm{W}}=\frac{1}{\sqrt{n}}(\ket{10\cdots 0}+\ket{01\cdots 0}+\cdots +\ket{00\cdots 1})$. Thus, it can distinguish some tripartite entanglement, at least the GHZ type from others. As discussed, $D_W=0$ if and only if $I=0$, implying no (geometric) entanglement in holography. Thus, it is evident from this that there is no GHZ-type tripartite entanglement in holography~\cite{Akers:2019gcv,Iizuka:2025bcc}.

Property (3) is very important as it indicates that the discord distinguishes entanglement and classical correlation, whose distinction may not be apparent from an exterior observer, who only has access to a coarse-grained picture~\cite{Verlinde:2020upt,Banerjee:2024fmh}. For example, $D_W(A|B)$ equals thermal entropy for a thermofield double (TFD) state dual to an eternal black hole while it becomes zero for the averaged TFD state called the thermo-mixed double (TMD) state, which is given by the diagonal part of the TFD density matrix. One might think it contradicts what we said earlier: $D_W=0$ if and only if $I=0$ in holography. However, we note that the TMD state usually appears in an effective picture where the state is coarse-grained over microstates. Thus, $D_W=0\Leftrightarrow I=0$ is still true with fine-grained entropies.

\paragraph{UV cutoff independence.---}
Unlike entanglement entropy, both classical and quantum correlations are UV-finite. This stems from the fact that both can be written as a relative entropy. Under the saddle-point approximation,
\begin{align}
	J_W(A|B)&=S(\rho_A^{dis} \!\parallel\! \rho_A),\\
	D_W(A|B)&=S(\rho_{AB}\!\parallel\! \rho_A\otimes\rho_B)-S(\rho_A^{dis}\!\parallel\! \rho_A),
\end{align}
where $\rho_A^{dis}$ denotes a holographic state of $A$ after the optimal measurement on $B$, corresponding to placing an end-of-the-world brane-like object along the boundary of the entanglement wedge of $B$. The relative entropy is UV-finite; moreover, it can be defined in general von Neumann algebras using the relative modular operator~\cite{araki1975inequalities,Witten:2018zxz}.

\section{Examples}\label{sec:ex}
\subsection{AdS$_3$/CFT$_2$}
As examples, let us explicitly work on $J_W(A|B)$ and $D_W(A|B)$ in AdS$_3$/CFT$_2$. For simplicity, the AdS radius is always rescaled to unity.

\paragraph{Poincar\'e AdS.---}
Parametrize each subregion as $A=[a_1,a_2]$ and $B=[b_1,b_2]$ in the Poincar\'e AdS$_3$. For convenience, we call the subregion in the middle by $C_1=[a_2,b_1]$.
The explicit forms of the holographic classical correlation and quantum discord between two intervals can be found as follows~\footnote{The convention of the UV cutoff $\epsilon$ is implicitly chosen such that the length of an empty subregion equals the UV cutoff (without a factor $2$), which is different from the convention used in~\cite{Kusuki:2022ozk}. If one follows their convention, $A,B$ becomes at least $2\epsilon$ while $C_1$ becomes at least $\epsilon/2$. This stems from the fact that the effective UV cutoff of the EWCS is different from the UV cutoff of entanglement entropy.}. When $I(A:B)=0$, $J_W(A|B)=D_W(A|B)=0$. When $I(A:B)>0$,
\begin{align}
	\frac{J_W(A|B)}{c/6}&=\max\qty(0,\log\qty[\frac{\abs{A}^2\abs{B}^2}{\abs{C_1}\abs{AC_1}\abs{C_1 B}\abs{AC_1 B}}]) \\
	\frac{D_W(A|B)}{c/6}&=\min\qty(\log\qty[\frac{\abs{A}^2\abs{B}^2}{\abs{C_1}^2\abs{AC_1 B}^2}], \log\qty[\frac{\abs{AC_1}\abs{C_1 B}}{\abs{C_1}\abs{AC_1 B}}])
\end{align}
where we used the Brown-Henneaux relation $\frac{3}{2G_N}=c$ to relate the Newton's constant to the central charge $c$. Here $\abs{A}=a_2-a_1$ denotes the length of interval $A$ on the boundary circle and likewise for other intervals.

~

\paragraph{Two-sided black hole.---}
Another interesting setup involves a black hole. Consider a TFD state
\begin{equation}
	\frac{1}{Z(\beta)}\sum_n e^{-\beta E_n/2} \ket{n}_{AC}\ket{n}_B, \quad Z(\beta)=\sum_n e^{-\beta E_n},
	\label{eq:TFD}
\end{equation}
which is dual to a two-sided black hole.
By measuring the black hole $B$, we evaluate the classical and quantum correlation. Since $J(A|B)$ takes the form of Holevo information, it tells us how much information we can get by measuring the black hole microstates. In contrast to previous studies~\cite{Bao:2017guc,Bao:2021ebo,Kudler-Flam:2021alo}, where typical black hole microstates were considered, the horizon has an atypical structure similar to the end-of-the-world brane in the current case. This is more akin to~\cite{Qi:2021oni}, in which the authors studied the Holevo information in the presence of the entanglement island. Their formula can be identified with ours through double holography~\cite{Almheiri:2019hni}. 

\begin{figure}
	\centering
	\includegraphics[width=\linewidth]{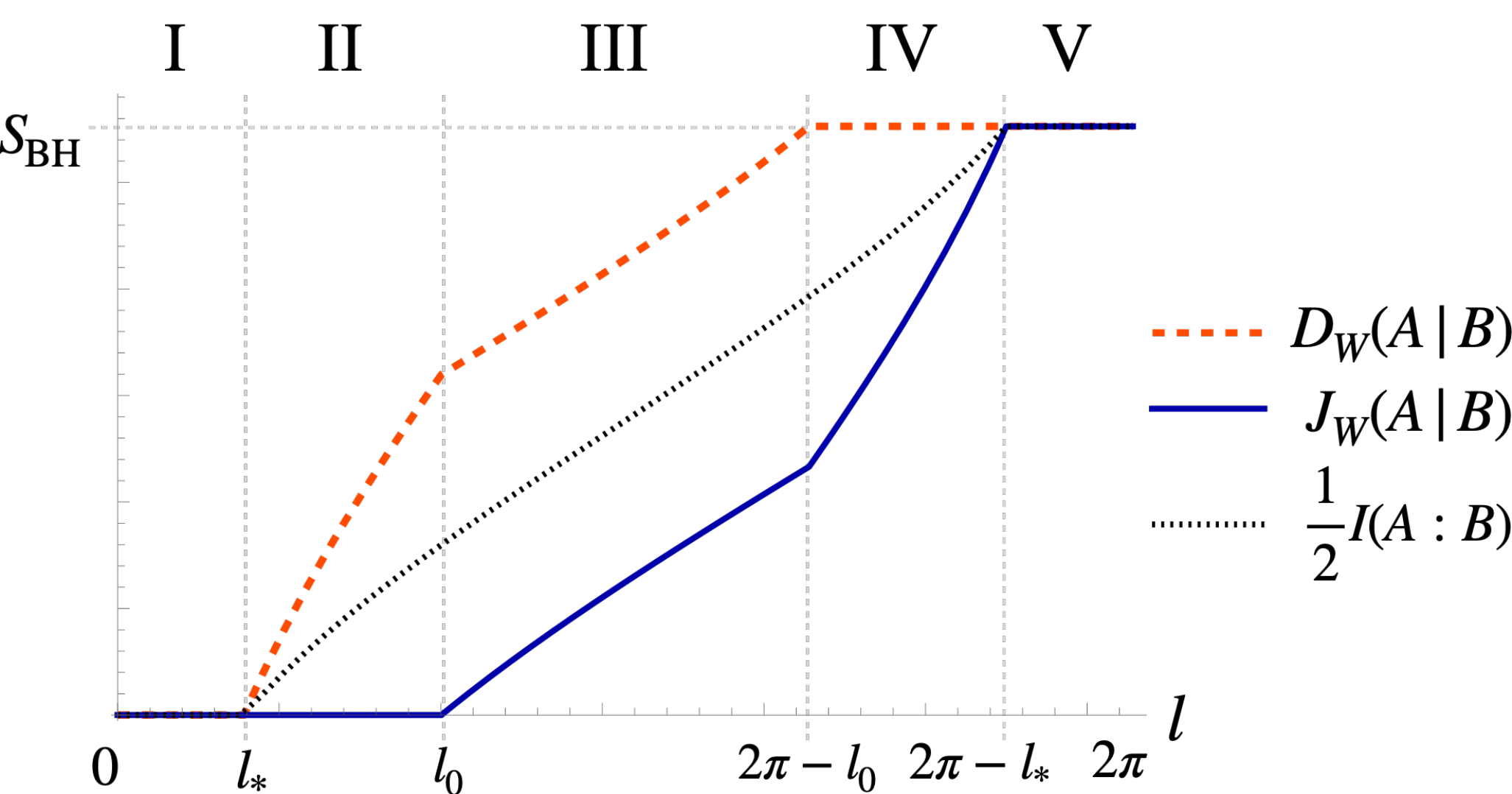}
	\caption{
    $J_W(A|B),D_W(A|B),I(A:B)/2$ of a TFD state~\eqref{eq:TFD} of temperature $0.14$ as we change the size of $A$ (denoted by $l$) from $0$ to $2\pi$. Depending on the value of $J_W$ and $D_W$, there are five distinct phases I--V. The definition of the phases and the threshold values are given in Appendix~\ref{app:BTZ}.
    }
	\label{fig:TFD}
\end{figure}

The explicit calculation is presented in Appendix~\ref{app:BTZ}.
The disentangled boundary leads to larger distinguishability, quantified by $J(A|B)$, when the temperature is large enough $\beta<\pi^2/\log(1+\sqrt{2})$. Fig.~\ref{fig:TFD} shows the plot of $J(A|B)$ and $D(A|B)$ together with $I(A:B)/2$ as we change the angle of $A=[0,l]$ from $l=0$ to $2\pi$.

From Fig.~\ref{fig:TFD}, we can explicitly confirm the properties discussed earlier. $J_W(A|B)$ and $D_W(A|B)$ become entanglement entropy $S_{\mathrm{BH}}$ in the pure state limit $l\rightarrow 2\pi$ and both are monotonically decreasing functions under the partial trace of $A$. $D_W(A|B)$ is vanishing only when $I(A:B)=0$ at leading order (geometrically separable). Furthermore, there is a hierarchy $J_W(A|B)\le E_{sq}(A:B)=I(A:B)/2\le D_W(A|B)$. In fact, this is also a generic property (at least in holography). We will come back to this point shortly.

~

\paragraph{One-sided black hole.---}
Let us consider another setup involving a black hole. A one-sided black hole is described by a Gibbs state $e^{-\beta H}/Z(\beta)$. By increasing the temperature $\beta^{-1}$, the state decoheres to a maximally mixed state, which has zero correlation. 

\begin{figure}
	\centering
	\includegraphics[width=0.85\linewidth]{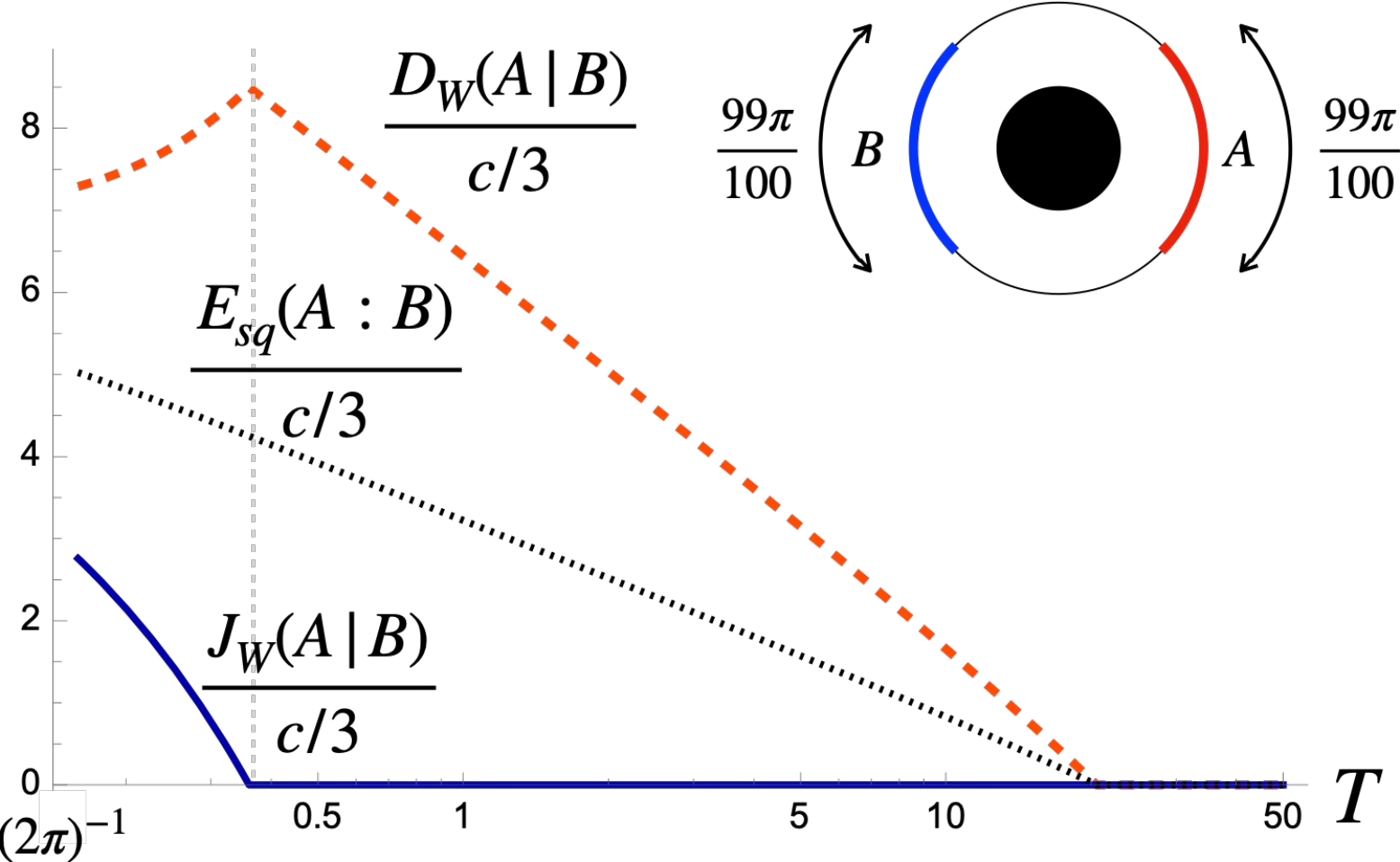}
	\caption{
    $D_W(A|B),E_{sq}(A:B),J_W(A|B)$ of a Gibbs state dual to a one-sided black hole as its temperature $T=\beta^{-1}$ increases (above the Hawking-Page threshold). They are rescaled by $c/3$ and depicted by red dashed, black dotted, and blue solid lines, respectively. The subsystems $A$ and $B$ are symmetrically placed as shown in the top right corner and each size is $99\pi/100$. 
    }
	\label{fig:BH-decoh}
\end{figure}

We take two subregions $A$ and $B$ symmetrically as shown in the top right corner of Fig.~\ref{fig:BH-decoh}. They are subsystems of a Gibbs state of temperature $T=\beta^{-1}$ in the dual CFT.
When the angle of each subsystem is sufficiently large, both the classical and quantum correlations are present to the leading order at zero temperature (vacuum). We examine how entanglement, classical correlation, and quantum discord change as the temperature increases. The detailed calculation is given in Appendix~\ref{app:hol-BH}. 

Fig.~\ref{fig:BH-decoh} shows the plot of the holographic correlation measures against temperature when the size of each subsystem is taken to be $49.95\%$ of the circumference $2\pi$. The plot shows the temperature above the Hawking-Page threshold $T_{HP}=(2\pi)^{-1}$ so the black hole phase is favored over the thermal AdS phase.

There are several interesting aspects of this plot. 
Firstly, while entanglement and classical correlation monotonically decrease as temperature increases, the holographic discord shows non-monotonic behavior since the classical correlation $J_W$ decays more rapidly than entanglement $E_{sq}$. In particular, the quantum discord increases as the temperature increases at the intermediate temperature---analogous to the anti-Unruh effect observed in two-qubit detectors~\cite{Henderson:2019uqo}. Although this may be counterintuitive as quantum discord typically arises from quantum coherence~\cite{Jin:2022vsa}, our result is consistent with a previous observation in two-qubit systems~\cite{Werlang_2010}. 

Secondly, as we increase the temperature, both holographic discord and entanglement cease to exist simultaneously for a sufficiently large but finite temperature. Naively, this seems different from previous observations in two-qubit or two-dimensional subspace coupled to an infinite dimensional system, where under some noise or for sufficiently high temperature/acceleration, entanglement experiences sudden death while discord remains finite~\cite{Werlang_2009,Werlang_2010,Hassan_2010,Chen-Yi-Xin_2010,Tian_2012,Datta_2009,Fuentes-Schuller:2004iaz,Celeri:2010xy,Landulfo:2009wg,Nambu:2011ae}.
However, this should be considered with caution. Since $J_W(A|B)$ is identified with the (one-way) distillable entanglement as noted in~\cite{Mori:2024gwe}, there is a sudden death of distillable entanglement prior to the death of discord even in this holographic setup. Thus, as previously suggested, discord is more robust to decoherence than a bipartite EPR-like entanglement. The difference between the distillable entanglement and squashed entanglement is difficult to highlight in previous studies as entanglement in $2\times2$ systems is always distillable~\cite{PhysRevLett.78.574,Horodecki:1998kf}.

To summarize, as we decohere the state by increasing the temperature, initially the quantum discord can increase during a rapid decrease of distillable entanglement. Once all the distillable entanglement shows the sudden death, the quantum discord decreases together with the squashed entanglement and they eventually experience sudden death at a sufficiently high temperature before reaching the infinite temperature.

\subsection{Haar random states}
Here we briefly present the asymptotic formulae for classical and quantum correlations of Haar random states. A mixed state $\rho_{AB}=\Tr_C \dyad{\Psi}_{ABC}$ obtained from a Haar random pure state $\ket{\Psi}_{ABC}$ has the following asymptotic formulae for $J(A|B)$ and $D(A|B)$. Let $n_A,n_B,n_C$ denote the number of qudits in each subsystem $A,B,C$ and the total number of qudits by $n$.
Despite their computation being very hard in general, in the limit of a large number of qudits with the ratio $n_{A,B,C}/n$ fixed, we can find the explicit formulae for $J(A|B)$ and $D(A|B)$ owing to the measure concentration~\cite{Hayden_2006}. They are given as
\begin{align}
	J(A|B)&=
	\begin{cases}
		0 \quad & (n_A<n_C)\\
		n_A-n_C \quad & (n_C \le n_A < n_B+n_C)\\
		n_B \quad & (n_B+n_C\le n_A) 
	\end{cases}
	,\\
	D(A|B)&=
	\begin{cases}
		I(A:B) \quad & (n_A+n_C>n_B,n_A<n_C)\\
		n_B \quad & (n_A+n_C>n_B,n_A>n_C)\\
		2n_A \quad & (n_A+n_C<n_B,n_A<n_C)\\
		n_A+n_C \quad & (n_A+n_C<n_B,n_A>n_C)
	\end{cases}
.
\label{eq:haar-discord}
\end{align}
Similar to holographic states, this can be explained from the geometric formula by identifying $E_F(A:C)=E_W(A:C)$. Here the EWCS is interpreted as the entanglement entropy of a pure state between $A$ and $C$ after measuring the rest of the qudits~\cite{Mori:2024gwe}. It is worth emphasizing that~\eqref{eq:haar-discord}, the asymptotic formula for the Haar random states, is not a conjecture but is rigorously proven.

\begin{figure}
    \centering
    \includegraphics[width=0.9\linewidth]{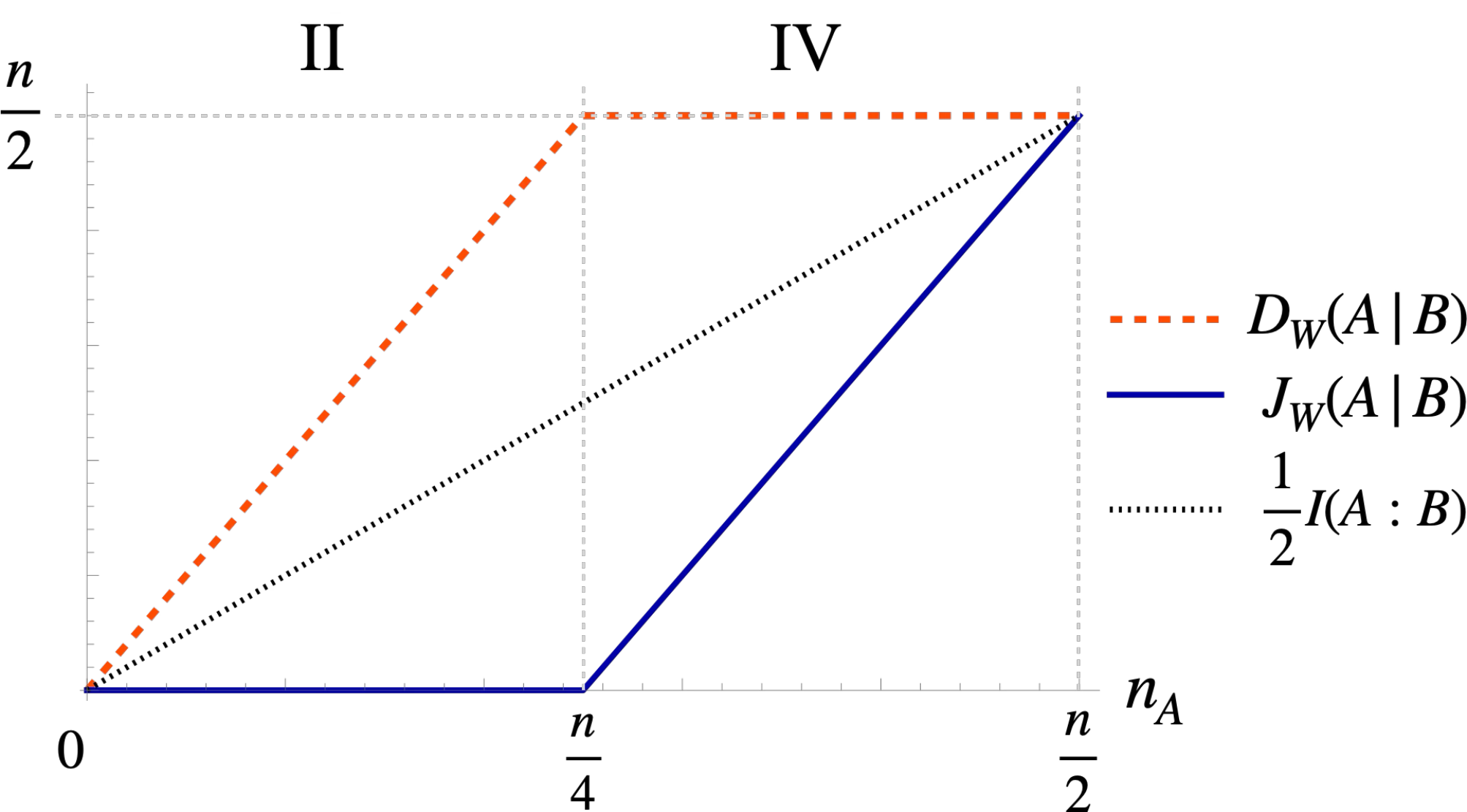}
    \caption{The classical and quantum correlations and a half of mutual information for $n$-qubit Haar random states in the large dimension limit. $n_A$ ranges from $0$ to $n/2$ while $n_B$ is fixed to be $n/2$. Corresponding to Fig.~\ref{fig:TFD}, there are two phases.}
    \label{fig:JD_Haar}
\end{figure}

Fig.~\ref{fig:JD_Haar} shows the leading order behavior of $J(A|B),D(A|B)$, and $I(A:B)/2$ as we vary $n_A$ from $0$ to $n/2$ while $n_B$ is fixed to be $n/2$. Similar to the holographic case (Fig.~\ref{fig:TFD}), we observe the same behavior despite Phase I, III, and V are absent since 
a Haar random state does not have an emergent bulk direction. However, note that if one takes an infinite temperature limit $T\rightarrow\infty$ and ignores the atypical phase (meaning $l_0=\pi$, in the holographic TFD setup), $l_\ast\rightarrow 0$ and Fig.~\ref{fig:JD_Haar} is correctly reproduced. This is consistent with a conventional lore that the infinite-temperature, typical black hole is similar to a Haar random state.

\section{Non-entanglement quantum correlation}\label{sec:NEQC}

\subsection{Quantum discord is larger than entanglement}
As the quantum discord captures quantum correlation other than entanglement, it is tempting to think it is greater than or equal to a faithful entanglement measure.
However, it remains unclear if there is such a quantitative hierarchy between quantum discord and entanglement in general. 
In fact, a related quantity called the geometric discord~\cite{PhysRevLett.105.190502}, defined through the Hilbert-Schmidt norm rather than the relative entropy, can become smaller than entanglement~\cite{Rana_2012}.

When the measurement in the definition of quantum discord is restricted to von Neumann's projective measurements, \cite{Piani_2012} showed that quantum correlation such as quantum discord is quantitatively larger than or equal to any measures of entanglement. However, when the discord is defined for general POVM, despite the optimal measurement being shown to be rank-one POVM~\cite{datta2008studies}, the optimal measurement is not necessarily projective, hence the quantitative hierarchical relation has not been proven yet~\footnote{For some specific cases, the quantum discord is observed to be greater than entanglement measures~\cite{qureshi2018hierarchy}. Furthermore, when there is no entanglement (known as the dissonance states~\cite{PhysRevLett.104.080501}), $D(A|B)>0$ is obviously greater than the entanglement, which is zero.}.

For holographic/Haar random states, this intuition is correct in the semiclassical/large dimension limit. Specifically, the quantum discord is greater than or equal to the squashed entanglement (up to subleading corrections):
\begin{equation}
	D(A|B)\ge E_{sq}(A:B). \label{eq:qd-ent}
\end{equation}

To show this, we first show that the classical correlation in these systems is less than or equal to the squashed entanglement:
\begin{equation}
	E_{sq}(A:B)\ge J(A|B).\label{eq:ent-cc}
\end{equation}
At an intuitive level, this is reasonable as the holographic and Haar random states are believed to have little classical correlation~\cite{Susskind:2014yaa, Nezami:2016zni,Hayden:2011ag,Dong:2021clv}. \eqref{eq:ent-cc} says the amount of classical correlation in these systems does not exceed the amount of entanglement, which is affirmative to the previous belief, and propels it to a more quantitative statement~\footnote{Note that the definition of the classical correlation employed here includes entanglement. For example, an EPR pair has $\log 2$ classical correlation. Thus, little classical correlation argued in the earlier literature does not necessarily mean $E_{sq}\gg J$.}.

We can show~\eqref{eq:ent-cc} in two ways -- holographically (geometrically) or operationally. Let us first begin with a holographic state. The conjecture $E_{sq}=I/2$ for holographic states \eqref{eq:hol-sq} leads
\begin{align}
	E_{sq}(A:B)-J_W(A|B)&= E_W(A:C)-\frac{I(A:C)}{2} \ge 0.
\end{align}
Note that $ABC$ forms a pure state and we used $E_W\ge I/2$ in the last inequality~\cite{Freedman:2016zud,Takayanagi:2017knl}. 

The conjecture $E_{sq}=I/2$ can be straightforwardly extended to the Haar random cases if we assume a geometric extension is optimal~\cite{Mori:2024gwe}. However, no rigorous proof is known yet. Nevertheless, without knowing the explicit form of the squashed entanglement, we can show~\eqref{eq:ent-cc} by utilizing the operational interpretation. 

As discussed in~\cite{Mori:2024gwe,Hayden_2006}, $J(A|B)$ is conjectured to lower bound the one-way distillable entanglement~\footnote{This is indeed true if $J\approx J_W$ at leading order.}. Since the squashed entanglement is lower bounded by the distillable entanglement~\cite{Christandl_2004}, $J(A|B)$ lower bounds $E_{sq}$. We stress that this neither depends on the conjecture $J=J_W$ nor $E_{sq}=I/2$. 
This proof basically stems from the construction of a one-way one-shot local operation and classical communication (LOCC) protocol to distill EPR pairs. This works for Haar random states and random (holographic) tensor networks. If we assume the existence of the disentangled basis in holographic conformal field theories, the same proof works for general holographic states.

Now, let us show that \eqref{eq:qd-ent} follows from \eqref{eq:ent-cc}. Recalling that $E_{sq}\le I/2$, it follows that
\begin{equation}
	D=I-J\ge I-E_{sq}\ge \frac{I}{2}\ge E_{sq},
\end{equation}
where the first inequality follows from the assumption \eqref{eq:ent-cc}. 

Note that \eqref{eq:ent-cc} does not generically hold (e.g. QC states). Hence, it remains an open problem whether \eqref{eq:qd-ent} is true for non-holographic/non-random states.

\subsection{Non-entanglement quantum correlation}
To further examine the nature of quantum correlation beyond entanglement, let us define \emph{non-entanglement quantum correlation} (NEQC) as a difference between the quantum discord and a faithful entanglement measure called the squashed entanglement: 
\begin{equation}
	\Delta Q(A|B) \equiv D(A|B)- E_{sq}(A:B).
\end{equation}
As shown earlier, this is non-negative at least for holographic and random states.

As shown below, we can immediately see that the NEQC is proportional to the Markov gap $h(A:C)$, which is dual to $2E_W(A:C)-I(A:C)$, in holography.
Using the conjectured duality $E_{sq}(A:B)=I(A:B)/2$, the NEQC is dual to the \emph{holographic NEQC}, defined as
\begin{equation}
	\Delta Q_W(A|B) = E_W(A:C) - \frac{I(A:C)}{2}=\frac{h(A:C)}{2}.
\end{equation}
Originally, the Markov gap is defined as the difference between the reflected entropy $S_R$, which will be defined explicitly in~\eqref{eq:ref-ent}, and the mutual information. This quantity is related to a certain Markov recovery problem necessarily involving the canonical purification~\cite{Hayden:2021gno}. This indicates that $h(A:C)$ is sensitive to the tripartite entanglement as seen from $AC$'s canonical purification~\cite{Zou:2020bly}.
Our formula indicates that there is an alternative interpretation in terms of the bipartite quantum correlation, namely, when $C$ purifies $AB$, a half of the Markov gap between $A$ and $C$ equals the NEQC between $A$ and $B$.

In AdS$_3$/CFT$_2$, the Markov gap is lower bounded by $\frac{c}{6}\log 2$ times the number of cross-section boundaries. In higher dimensions, although there is no constant lower bound, the Markov gap is still $O(1/G_N)$ if and only if the entanglement wedges are connected~\cite{Hayden:2021gno,Takayanagi:2017knl}. This suggests that a large amount $\sim O(c)$ of NEQC between $A$ and $B$ exists in holography whenever $A$ and the environment $C$ have a connected entanglement wedge.

This is expected because the Markov gap is proposed as a quantity relevant to the tripartite entanglement and the tripartite entanglement implies a departure from purely bipartite entanglement. But as noted in~\cite{Akers:2019gcv,Iizuka:2025ioc}, the Markov gap tells us more. The GHZ entanglement is one of the tripartite entanglement, however, it has a vanishing Markov gap. This is evident from our alternative interpretation of the Markov gap --- the GHZ state gives rise to a classical correlation for any sub-parties so it does not offer a quantum correlation among them. Thus, our interpretation of the Markov gap clarifies its meaning --- a nonvanishing Markov gap implies more than tripartite entanglement. 

This connection to the tripartite entanglement can be seen from the alternative interpretation of the classical correlation as the (one-shot) one-way distillable entanglement. In~\cite{Mori:2024gwe}, we defined $J_W(A:B)=\max(J_W(A|B),J_W(B|A))$ and argued it is dual to the one-way distillable entanglement in either direction. This motivates us to define a symmetrized version of the holographic quantum discord, $D_W(A:B)=\min(D_W(A|B),D_W(B|A))$, from which the holographic symmetrized NEQC is defined as
\begin{equation}
	\Delta Q_W(A:B) = \frac{\min(h(A:C),h(B:C))}{2}.
\end{equation}
Alternatively, by using the identification $E_D=J_W$, we have
\begin{align}
	\Delta Q_W(A:B) &= \frac{I(A:B)}{2} - E_D(A:B)\\
	&= E_{sq}(A:B)-E_D(A:B).\label{eq:NEQC-undist}
\end{align}
This indicates that the holographic NEQC is also dual to the amount of undistillable entanglement. 
Without symmetrizing, $\Delta Q_W(A|B)$ is dual to the \emph{one-way} undistillable entanglement from $B$ to $A$.
Since $E_D$ focuses on the bipartite entanglement distillation, $\Delta Q_W>0$ implies the existence of the tripartite entanglement~\footnote{Recently, a related paper appeared where the authors discuss a connection between non-distillability and nonzero Markov gap~\cite{Jin:2025zmt}. However, note that the Markov gap appearing in this paper is between a subsystem and the purifying partner while the Markov gap in their paper is between two subsystems of interest.}.

We also note that the symmetrized discord is monotone under local operations because of property (6) of the quantum discord. This is also natural from the above relation with undistillable entanglement because entanglement is LOCC monotone. 
There have been some debates on whether the quantum discord is really a correlation measure or not since the (unsymmetrized) quantum discord is not LO monotone, stemming from the fact that the quantum coherence can be created locally. The symmetrized version makes it manifestly LO monotone so it is better behaved as a correlation measure although it remains open whether this symmetrized discord is LOCC monotone or not. 

The Markov gap is originally defined in terms of the reflected entropy, which is dual to twice the EWCS. This motivates us to study a boundary dual of the holographic discord even outside of holographic regimes, which we will explore in the next section.

\section{Boundary dual}\label{sec:reflected}
Having explored the bulk duals of classical and quantum correlations, we now consider their alternative boundary counterparts toward optimization-free measures applicable to generic quantum systems.

\subsection{Reflected correlation measures}
In AdS/CFT, a single geometric quantity can have multiple boundary duals. While the duality of the EWCS with the entanglement of formation leads to $D=D_W$, let us now consider its duality with the reflected entropy, defined as~\cite{Dutta:2019gen}
\begin{equation}
	S_R(A:B)=S(\rho_{AA^\ast}),
    \label{eq:ref-ent}
\end{equation}
where $\rho_{AA^\ast}=\Tr_{BB^\ast}\dyad*{\rho^{1/2}}$ and $\ket*{\rho^{1/2}}_{AA^\ast BB^\ast}$ is the canonical purification of $\rho_{AB}$.
Among various boundary duals of the EWCS, the reflected entropy does not require any optimizations and is well-defined in any quantum systems.


Because both classical correlation $J$ and discord $D$ require a difficult optimization~\cite{Huang_2014}, we now define \emph{reflected} versions that involve no optimization but capture the same qualitative behavior. Let us define \emph{reflected classical correlation} $J_R$, \emph{reflected discord} $D_R$, and \emph{reflected NEQC}  $\Delta Q_R$ by replacing the EWCS with $S_R/2$:
\begin{equation}
	\begin{split}
		J_R(A|B)&\equiv S_A - \frac{S_R(A:C)}{2},\\
		D_R(A|B)&\equiv S_B -S_{AB}+\frac{S_R(A:C)}{2},\\
		\Delta Q_R (A|B) &\equiv \frac{h(A:C)}{2}\equiv \frac{S_R(A:C)-I(A:C)}{2},
	\end{split}
\label{eq:bdy-dual}
\end{equation}
where $ABC$ forms a pure state.
For holographic and random states, they are dual to $J_W,D_W,\Delta Q_W$, and hence they are also dual to the original measures $J,D,\Delta Q$. However, for generic states, the reflected measures can differ from the original measures. However, they satisfy several desirable properties:
\begin{enumerate}
	\item They do not involve any optimizations.
	\item They are invariant under local unitary transformations.
	\item $0\le J_R(A|B),D_R(A|B) \le I(A:B)$.
	\item $J_R(A|B)=D_R(A|B)=S_A=S_B$ for pure $AB$.
	\item $J_R(A|B)\le \min(S_A,S_B)$.
	\item $D_R(A|B)\le S_B$.
	\item $\Delta Q_R(A|B)\ge 0$.
	\item $\Delta Q_R(A|B)=0$ when $\rho_{AB}$ is either a pure state or a locally distinguishable QC state. We expect this because a correlation in a pure state is only entanglement.
\end{enumerate}
Property (1) directly follows from their definitions. Property (2) is evident as they are functions of the von Neumann entropies. To show property (3), it is sufficient to show $J_R, D_R\ge 0$. $J\ge 0$ follows from $S_R(A:B)=S_{AA^\ast}\le 2S_A$. $D_R\ge 0$ follows from 
\begin{equation}
	D_R(A|B)\ge \frac{I(A:B)}{2}\ge 0,
\end{equation}
where the first inequality follows from $S_R\ge I$~\cite{Dutta:2019gen}.
Property (4) follows immediately by substituting $C=\emptyset$. For property (5), $J_R\le S_A$ is obvious from the non-negativity of $S_R$. $J_R\le S_B$ follows from $S_B-J_R(A|B)=S_B-S_A+\frac{S_R(A:C)}{2}=\frac{I(B:C)}{2}\ge 0$. Property (6) follows from $S_R(A:C)\le 2S_C=2S_{AB}$. 
Property (7) follows from the non-negativity of the Markov gap (equivalently, $S_R\ge I$).
For property (8), when $\rho_{AB}$ is a pure state, from property (4), $D_R=J_R\Rightarrow \Delta Q_R=0$. When $\rho_{AB}$ is quantum-classical~\eqref{eq:qc-state}, 
its purification becomes
\begin{equation}
	\ket{\Psi}_{ABC}=\sum_i \sqrt{p_i} \ket*{\rho_i^{1/2}}_{AC}\ket{ii}_{BC}
\end{equation}
and the reduced density matrix on $AC$ is given by
\begin{equation}
	\rho_{AC}=\sum_i p_i \dyad*{\rho_i^{1/2}}_{AC}\otimes \dyad{i}_C.
\end{equation}
From its canonical purification
\begin{equation}
	\ket*{\rho^{1/2}}_{ACA^\ast C^\ast}=\sum_i \sqrt{p_i} \ket*{\rho^{1/2}}_{AC} \ket*{\rho^{1/2}}_{A^\ast C^\ast} \ket{ii}_{CC^\ast},
\end{equation}
one can compute the reflected entropy and find
\begin{equation}
	\Delta Q_R (A|B) = 
	\frac{S\qty(\sum_i p_i \rho_i^{\otimes 2})-S\qty(\sum_i p_i\rho_i)-\sum_i p_i S(\rho_i)}{2}.
\end{equation}
While this does not vanish in general, if $\rho_{AB}$ is locally distinguishable, i.e. $\rho_i \perp \rho_j$ for $i\neq j$, $\Delta Q_R(A|B)$ becomes vanishing.

Unlike the quantum discord, the reflected discord does not vanish even if $\rho_{AB}$ is quantum-classical or even classically correlated. Nevertheless, $\Delta Q_R$ vanishes when it is quantum-classical and locally distinguishable. Although we need this additional condition of the local distinguishability, the class of states with vanishing $\Delta Q_R$ is still larger than the set of classically correlated states and pure states.

From these observations, we expect $D_R$ upper bounds $D$ and $\Delta Q_R$ might be even more close to the original measure $\Delta Q$. In fact, $D_R\ge D$ holds generically. Its proof is presented in~\cite{Louisia:2025bxz}.

Unfortunately, comparing $\Delta Q_R$ with $\Delta Q$ is very difficult as the latter involves not only the computation of the quantum discord but also the squashed entanglement, whose analytic form is barely known. It is tempting to make the comparison in some limiting cases such as when $E_D=E_{sq}=E_C$~\cite{Moharramipour:2024hnu}.

\subsection{Two-qubit example}
Let us study these reflected measures in a non-holographic system. Since a comparative study with the original classical and quantum discord is necessary for our purpose, we focus on a one-parameter family of rank-two states, where an analytical formula of quantum discord is known.

Let us consider a rank-two two-qubit state of the following form:
\begin{equation}
	\rho_{AB}(\theta) = \frac{\dyad{0}_A\otimes\dyad{0}_B+\dyad{1}_A\otimes\dyad{\theta}_B}{2},
	\label{eq:param-state}
\end{equation}
where
\begin{equation}
	\ket{\theta}=\cos\theta \ket{0}+\sin\theta \ket{1}.
\end{equation}
$\rho_{AB}(\theta)$ interpolates a classically uncorrelated state ($\theta=0$), a discordant state ($\theta=\pi/4$), and a classically maximally correlated state in the two-dimensional subspace ($\theta=\pi/2$).

Thanks to the Koashi-Winter relation, the analytic expression of the quantum discord of rank-two states is known~\cite{Shi_2011}.
The classical correlation and the quantum discord for~\eqref{eq:param-state} are
\begin{align}
	J(A|B)&=\log 2 -H\qty(\sin^2\qty(\frac{\theta}{2}-\frac{\pi}{4})), \\
	D(A|B)&=H\qty(\sin^2\frac{\theta}{2})+H\qty(\sin^2\qty(\frac{\theta}{2}-\frac{\pi}{4})) -\log 2,
\end{align}
where $H(p)=-p\log p - (1-p)\log(1-p)$ is the binary entropy.
The NEQC $\Delta Q$ equals the quantum discord as the state is always separable.

The reflected correlation measures are
\begin{align}
	J_R(A|B)&=\frac{1}{2} \qty(\log 2 -  H\qty(\sin^2\qty(\frac{\theta}{2}-\frac{\pi}{4}))), \\
	D_R(A|B) & =  H\qty(\sin^2\frac{\theta}{2}) + \frac{H\qty(\sin^2\qty(\frac{\theta}{2}-\frac{\pi}{4}))-\log 2}{2} ,\\
	\Delta Q_R(A|B) & =  \frac{H\qty(\sin^2\frac{\theta}{2}) +  H\qty(\sin^2\qty(\frac{\theta}{2}-\frac{\pi}{4})) - \log 2}{2},
\end{align}
where $H(p)$ is a binary entropy function.
The detail of these calculations is given in Appendix~\ref{app:rank2}.

The reflected correlation measures are related to the original measures as follows.
\begin{align}
	J(A|B) & = 2 J_R(A|B)\\
	D(A|B) & = D_R(A|B) - J_R(A|B)\\
	\Delta Q(A|B) & = 2 \Delta Q_R(A|B).
\end{align} 
Surprisingly, two reflected measures, $J_R$ and $\Delta Q$, match with the original optimized measures up to a constant, even though the states in this case are far away from the holographic regime -- two-dimensional and separable.
As speculated in the previous subsection, we see that $\Delta Q_R$ is more similar to the original quantity than the quantum discords. 

It is worth noting that, in this case, the reflected discord surpasses the diagonal discord, another measure proposed earlier to estimate the quantum discord~\cite{Liu_2019}. It is defined as
\begin{align}
	D_{\rm diag}(A|B)&=I(A:B)-J_{\rm diag}\\
    &=S_B-S_{AB}+\sum_i p_i S(\rho_i^A),
\end{align}
where the measurement is chosen to be a projective measurement by the eigenbasis of $\rho_B$. Explicitly,
\begin{equation}
	\rho_i^A = \frac{1}{p_i}\Tr_B((\bm{1}_A\otimes \Pi_B^i)\rho_{AB}),
\end{equation}
where $\Pi^i = \dyad{\nu_i}$ such that $\rho_B=\sum_i \lambda_i \dyad{\nu_i}$. $\lambda_i$ is the eigenvalues of $\rho_B$ and $p_i=\Tr(\Pi_B^i\rho_{AB})$.

Because of a specific choice of measurement, the diagonal discord upper bounds the quantum discord. The diagonal classical correlation for~\eqref{eq:param-state} is computed as
\begin{equation}
	J_{\rm diag}(A|B)=0.
\end{equation}
Thus, the diagonal discord only captures the mutual information part while the reflected measures capture the other $\theta$ dependence as well as the constant term proportional to $\log 2$.

As one can see from the diagonal discord, finding an optimal measurement basis is not a trivial task. For instance, one may attempt to estimate the classical correlation by a random measurement $\{2U\dyad{0}U^\dag\}_U$ on $B$~\footnote{The factor $2$ is added for the normalization in accord with the Haar measure: $\int dU 2U\dyad{0}U^\dag = \mathbf{1}$.}.
Denoting the corresponding probability by $p_U$, the classical correlation evaluated with the random measurement basis after the Haar random averaging is given by
\begin{equation}
    \overline{J_{\rm rand}(A|B)}=\log 2-\int\dd U\, p_U H\qty(q_U(\theta))
    \label{eq:j-rand-1}
\end{equation}
where
\begin{equation}
    p_U = \abs{\!\ev{U}{0}}^2+\abs{\!\mel{\theta}{U}{0}}^2
\end{equation}
and
\begin{equation}
    q_U(\theta) = \frac{\abs{\!\ev{U}{0}}^2}{\abs{\!\ev{U}{0}}^2+\abs{\!\mel{\theta}{U}{0}}^2}.
\end{equation}

\begin{figure}
    \centering
    \includegraphics[width=1\linewidth]{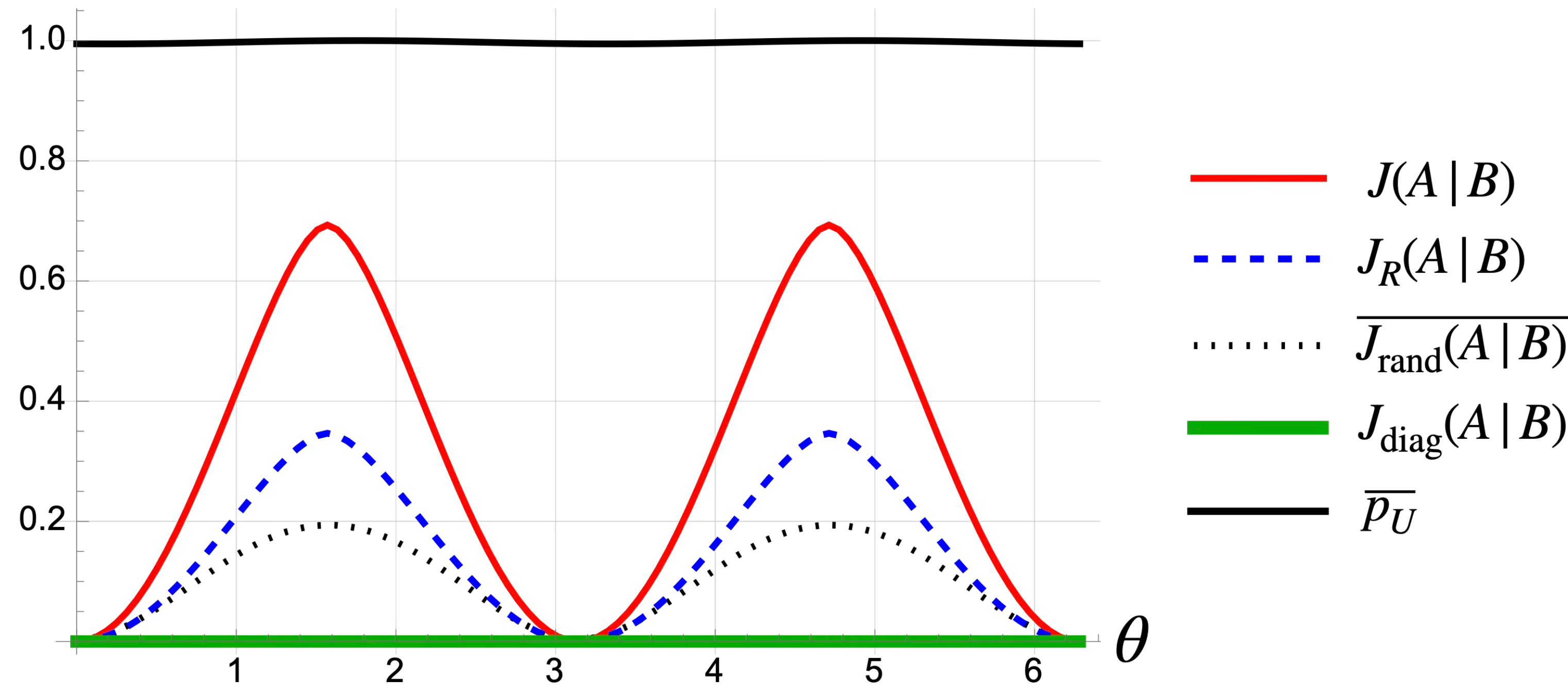}
    \caption{The classical correlation measures for the rank-two qubit states~\eqref{eq:param-state} with $0\le \theta\le 2\pi$. The solid red line denotes the true classical correlation $J(A|B)$ and the blue dashed line is the reflected one $J_R(A|B)$. The dotted black line is $\overline{J_{\rm rand}}(A|B)$, the averaged classical correlation evaluated with the random basis. The Haar random average is taken over 10,000 samples. The black line denoting the averaged $p_U$ is approximately unity, ensuring the normalization of the random basis measurement. The thick green line denoting $J_{\rm diag}(A|B)$ is zero for any $\theta$.}
    \label{fig:rank-2}
\end{figure}

The various classical correlations are summarized in Fig.~\ref{fig:rank-2}. The Haar random average is taken over 10,000 samples and one can confirm from the averaged probability $\overline{p_U}=\int dU\, p_U$ being approximately unity that $\{U\dyad{0}U^\dag\}_U$ forms the measurement basis after averaging.
It is clear from the plot that $J_R$ is the largest among other classical correlations with different choices of measurements. $J$ and $J_R$ are the same up to the overall factor while $\overline{J_{\rm rand}}$ is not and smaller than these two measures. The diagonal classical correlation is zero, indicating the eigenbasis is not optimal at all. 

\section{Multipartite generalization}\label{sec:multi}
The relation between the NEQC and Markov gap or undistillable entanglement indicates the existence of tripartite entanglement, its characterization remains unclear.

We now propose five distinct strategies for extending classical correlation $J$ and discord $D$ to $n$ parties. Each approach replaces the bipartite mutual information or entanglement entropy with a different $n$-partite quantity: (i) total correlation, (ii) dual total correlation, (iii) sequentially measured mutual information, (iv) $n$-partite information, and (v) holographic multi-entropy. We describe each in turn.


\subsection{Total correlation}
Since both the classical correlation and quantum discord are based on mutual information, various generalizations can be obtained by generalizing the mutual information to a multipartite setting. Recalling the (bipartite) mutual information can be written as the relative entropy
\begin{equation}
    I(A:B)_\rho = S(\rho_{AB} \!\parallel\! \rho_A\otimes\rho_B),
\end{equation}
a natural generalization to $n$ partitions is given by the so-called total correlation (TC)~\cite{Groisman_2005,5392532}:
\begin{align}
    I^{TC}(A_1:\cdots : A_n) &=  S(\rho_{A_1\cdots A_n} \!\parallel\! {\textstyle\bigotimes_{k=1}^n}\rho_{A_k}), \\
    & = \sum_{k=1}^n S(A_k) - S(A_1\cdots A_n) \\
    & = \sum_{k=1}^{n-1} S(A_k) - S(A_1\cdots A_{n-1}|A_n).
\end{align}
As one can see from the last line, the total correlation is defined as a sum of individual uncertainties minus the conditional joint uncertainty. When $n=2$, it reduces to the standard mutual information $I(A_1:A_2)$.

We know the bipartite classical correlation $J(A|B)$ is obtained by replacing the conditioning in the mutual information with the classical conditioning by a local measurement and the average over the outcome. Similarly, we define the \emph{TC-classical correlation} by
\begin{align}
    &J^{TC}(A_1 : \cdots :\! A_{n-1}|A_n) \nonumber\\
    =& \max\qty( \sum_{k=1}^{n-1} S(A_k) - \overline{S(A_1\cdots A_{n-1})_{A_n}}).
\end{align}
$S(B)_C$ denotes the von Neumann entropy of $B$ after the measurement on $C$ and the bar represents the average over the measurement outcome, weighted by the measurement probability. The maximization is taken over all possible POVM on $A_n$. 
From the definition, it is clear that $J^{TC}(A|B)=J(A|B)$.

Based on this, we define the \emph{TC-quantum discord} by
\begin{align}
    &\phantom{\,=}D^{TC}(A_1:\cdots :\! A_{n-1}|A_n) \nonumber \\
    &= I^{TC}(A_1:\cdots :A_n) - J^{TC}(A_1 : \cdots :\! A_{n-1}|A_n) \nonumber\\
    & = \min \overline{S(A_1\cdots A_{n-1})_{A_n}} - S(A_1\cdots A_{n-1}|A_n).
\end{align}
It can be seen by comparing to the definition of the bipartite quantum discord~\eqref{eq:q-discord} that this multipartite generalization in fact reduces to the bipartite measure:
\begin{equation}
    D^{TC}(A_1:\cdots :\! A_{n-1}|A_n) = D(A_1\cdots A_{n-1}|A_n)
\end{equation}
although the TC-classical correlation differs from the bipartite one.

Similar to the bipartite classical correlation and quantum discord, the TC measures are UV finite and the TC-classical correlation is manifestly non-negative (as well as the TC-quantum discord). This follows from the concavity of the von Neumann entropy and non-negativity of the $(n-1)$-partite TC, which can be shown from the non-negativity of the relative entropy.

\subsection{Dual total correlation}
Another generalization of the mutual information is known as the dual total correlation (DTC), defined as~\cite{Kumar_2017,HAN1978133,Lee:2023kgy,Kumar:2023mkx}
\begin{align}
    I^{DTC}(A_1:\cdots:\! A_n) &= S(A) - \sum_{k=1}^n S(A_k|A\setminus A_k) \\
    & = \sum_{k=1}^n S(A\setminus A_k) -(n-1) S(A), \nonumber
\end{align}
where $A=\bigcup_{k=1}^n A_k$. Again, it reduces to $I(A:B)$ when $n=2$.

By replacing the quantum conditioning with the classical conditioning by a measurement, we define the \emph{DTC-classical correlation} as
\begin{equation}
    J^{DTC}(A_1:\cdots :\!A_n) = \max\qty( S(A) - \sum_{k=1}^n \overline{S(A_k)_{A\setminus A_k}} ),
\end{equation}
where $A=\bigcup_{k=1}^n A_k$ as above.
The maximization is taken over all possible measurements in the second term.

We define the \emph{DTC-quantum discord} as
\begin{align}
    &\phantom{==}D^{DTC}(A_1:\cdots :\! A_n) \nonumber\\
    & = I^{DTC}(A_1:\cdots:\! A_n) - J^{DTC}(A_1:\cdots :\!A_n)\nonumber\\
    & = \min \sum_{k=1}^n \overline{S(A_k)_{A\setminus A_k}} - \sum_{k=1}^n S(A_k|A\setminus A_k).
\end{align}

Unlike other classical/quantum correlation measures, it is symmetric under the permutation of $A_k$'s. The downside of this quantity is that it does not reduce to the bipartite classical correlation/quantum discord when $n=2$ although it is related to the bipartite measures as
\begin{align}
    J^{DTC}(A:B)&=J(A|B)+J(B|A)-I(A:B),\\
    D^{DTC}(A:B)&=D(A|B)+D(B|A).
\end{align}

\subsection{TC after sequential measurements}
In~\cite{Radhakrishnan_2020}, the authors propose a multipartite generalization of quantum discord based on a sequential measurement $A_1\rightarrow A_2 \rightarrow \cdots \rightarrow A_n$. The arrow represents the feedforward of the measurement results. Eventually, the measurement is applied to at most $(n-1)$ parties. To derive the multipartite generalization of the correlation measures, first note that the TC can be written as a sum of the mutual information:
\begin{equation}
    I^{TC}(A_1:\cdots :\! A_n) = \sum_{k=1}^{n-1} I(A_1\cdots A_k: A_{k+1}).
\end{equation}
Since each mutual information is decomposed into the entropy and the quantum conditional entropy, we can obtain the classical mutual information by replacing the conditioning by the classical one with a measurement to obtain the classical correlation. Now, the multipartite generalization of the classical correlation under the sequential measurement is defined as
\begin{equation}
    J^{seq}(A_1\rightarrow \cdots \rightarrow A_n) =\max \sum_{k=2}^n (S(A_k) - \overline{S(A_k)_{A_1\cdots A_{k-1}}}),
\end{equation}
where the maximization is taken over all possible sequential POVM.
The multipartite generalization of the quantum discord is defined as
\begin{align}
    &\phantom{==} D^{seq}(A_1\rightarrow \cdots \rightarrow A_n) \nonumber\\
    &=  I^{TC}(A_1:\cdots :\! A_n) - J^{seq}(A_1\rightarrow \cdots \rightarrow A_n) \nonumber\\
    &= \min \sum_{k=1}^n S(A_k)_{A_1\cdots A_{k-1}} - S(A_1\cdots A_n).
\end{align}
Here we defined the $k=1$ term of the first term by $S(A_1)$. 

When $n=2$, these correlation measures reduce to their bipartite versions. Namely, $J(A_1\rightarrow A_2)=J(A_2|A_1)$ and $D(A_1\rightarrow A_2)=D(A_2|A_1)$. This generalization depends on the ordering of the subsystems by sequentially conditioning the entropies.

\subsection{$n$-partite information}
Another possible generalization is given by the $n$-partite information~\cite{1057469,79902},
\begin{equation}
    I_n(A_1:\cdots :\! A_n) = - \sum_{\emptyset\neq J\subseteq \{1,\cdots,n\}} (-1)^{\abs{J}} S\qty({\textstyle\bigcup_{j\in J}} A_j).
\end{equation}
This quantity, especially when $n=3$, has been intensively studied in holography and quantum field theory~\cite{Casini:2008wt,Hayden:2011ag,Bao:2015bfa,Mirabi:2016elb,Agon:2021lus,Agon:2022efa,Agon:2024zae}.

One particularly nice feature of $I_n$ is that there is a recursive relation with the $(n-1)$-partite conditional mutual information, similar to the mutual information $I(A:B)=S(A)-S(A|B)$:
\begin{equation}
    I_{n}(A_1:\cdots:\!A_{n})=I_{n-1}-I(A_1:\cdots:\!A_{n-1}|A_{n}),
\end{equation}
where the $n$-partite conditional mutual information is defined as
\begin{equation}
    I(A_1:\cdots:\!A_n|A_{n+1})=-\!\!\!\!\!\!\!\sum_{J\subseteq \{1,\cdots,n\}} \!\!\!\!\!(-1)^{\abs{J}} S\qty({\textstyle\bigcup_{j\in J}} A_j \cup A_{n+1}).
\end{equation}
Note that the summation here includes $J=\emptyset$.

This observation motivates us to define the \emph{$n$-partite classical correlation} by replacing the quantum conditioning with the classical one with a measurement~\footnote{The tripartite classical correlation is called the accessible tripartite information in~\cite{LoMonaco:2023xws}.}:
\begin{equation}
    J_n(A_1:\cdots:\!A_n)=\max(I_{n-1} - \overline{I(A_1:\cdots:\!A_{n-1})_{A_n}}),
\end{equation}
where the maximization is taken over all POVMs on $A_n$.

Then, it is straightforward to define the \emph{$n$-partite quantum discord} by 
\begin{align}
    &\phantom{==}D_n(A_1:\cdots:\!A_n)\nonumber\\
    &=\min\overline{I(A_1:\cdots:\!A_{n-1})_{A_n}}-I(A_1:\cdots:\!A_{n-1}|A_{n})
\end{align}

While the $n$-partite classical correlation and quantum discord are UV finite, their sign is indefinite as so is $I_n$~\cite{Hayden:2011ag}. This makes them difficult to interpret as some informational quantity per se.

\subsection{Holographic multi-entropy}
Finally, let us propose a new generalization of correlation measures based on the holographic proposals of multipartite entanglement. If one goes back to the holographic formula of the classical correlation~\eqref{eq:cc-hol}, $J(A|B)$ is defined by the entropy, which is proportional to the area of the minimal surface \emph{homologous} to $A$ minus the post-measurement entropy, whose surface is still homologous to $A$ but can anchor on the boundary of the entanglement wedge of $B^c$. This comes from the fact that the boundary of the entanglement wedge becomes disentangled after a suitable measurement in $B$~\cite{Mori:2024gwe}.

\begin{figure}
    \centering
    \includegraphics[width=0.35\linewidth]{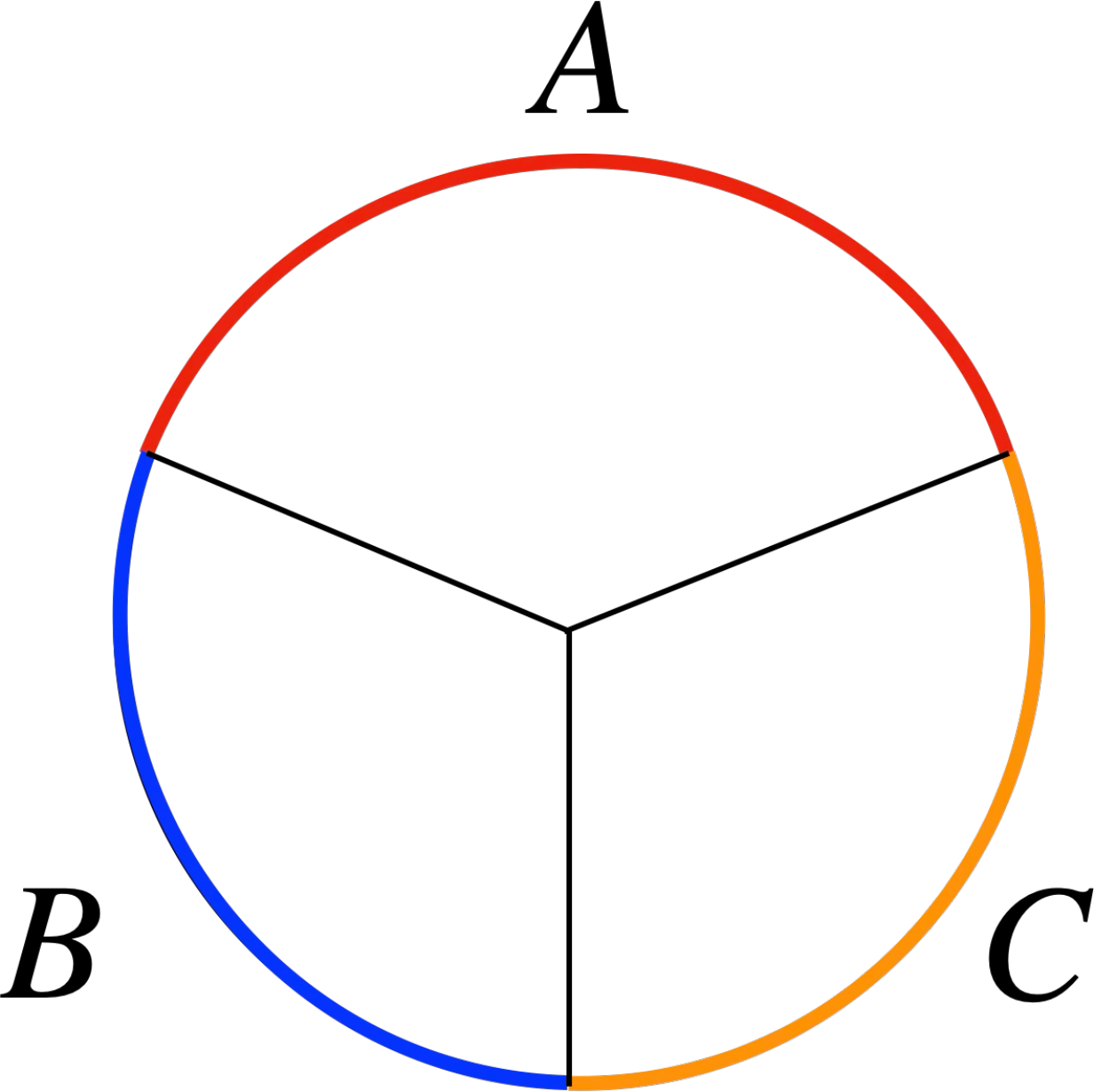}
    \caption{The bulk dual of the multi-entropy $S(A:B:C)$ in AdS$_3$/CFT$_2$. It is defined as the total area of the minimal surfaces that meet at a single bulk point.}
    \label{fig:mult-ent}
\end{figure}

From this observation, one possibility for generalizing $J(A|B)$ to multipartite settings such as $J(A:B|C)$ is to replace the entropy with the multi-entropy $S(A:B:C)$~\cite{Gadde:2022cqi,Gadde:2023zzj,Penington:2022dhr,Harper:2024ker,Gadde:2024taa}. Its putative bulk dual is illustrated in Fig.~\ref{fig:mult-ent}. For a pure, tripartite holographic state, it is given by the area of a minimal tri-way cut divided by $4G_N$. The minimal tripartition is chosen to be homologous to each subregion and each cut meets at a single bulk point, which is being optimized.
In general, we can define \emph{holographic multi-entropy} for $n$ boundary subsystems $\{A_1,A_2,\cdots,A_n\}$ by
\begin{equation}
    S(A_1:A_2:\cdots:A_n)=\min_{\mathcal{W}} \frac{\mathrm{Area}(\mathcal{W})}{4G_N}
    \label{eq:mult-ent-def}
\end{equation}
where minimization is taken over a collection of surfaces (brane web) $\mathcal{W}$ that anchors at the boundaries of all subregions $\{A_1,A_2,\cdots,A_n\}$ and includes sub-webs satisfying the homology condition to all the subregions.

An immediate obstacle to our generalization is that the multi-entropy is usually defined with a pure multipartite state.
While the definition~\eqref{eq:mult-ent-def} can naively be extended to mixed states, there remains a subtlety regarding the boundaries adjacent to a purifying system.
One existing bulk generalization to a mixed-state scenario is the multipartite EWCS $E_W(A:B:C)$~\cite{Yuan:2024yfg,Bao:2018gck}. While this quantity reduces to the original multi-entropy in the pure state limit, it resembles more with the EWCS. Indeed, if one takes a bipartite mixed-state limit, it becomes the EWCS rather than the entanglement entropy. Thus, a naive substitution will not extrapolate the bipartite case.

\begin{figure}
    \centering
    \includegraphics[width=0.85\linewidth]{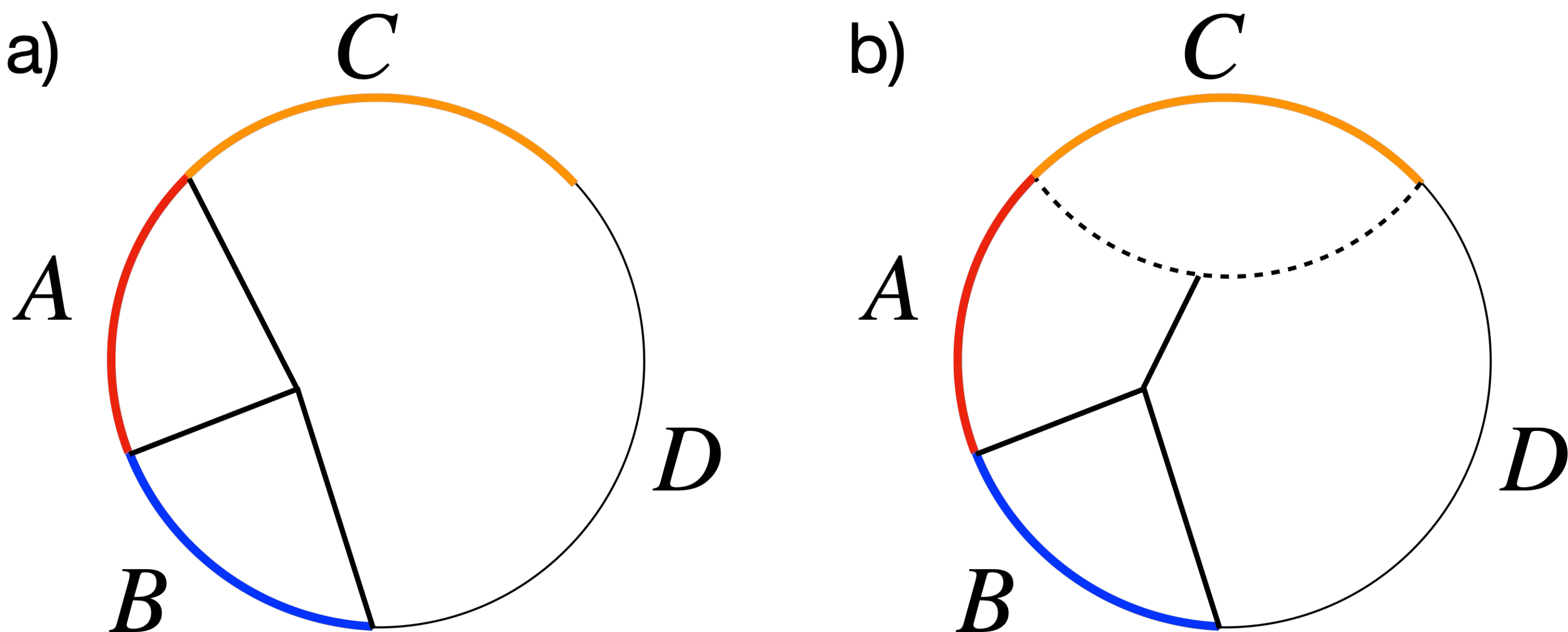}
    \caption{The mixed-state holographic multi-entropy $S(A:B;C)$ of $\rho_{ABC}$ a) before and b) after measuring $C$. $D$ denotes the purifying boundary subregion. 
    }
    \label{fig:multi-J}
\end{figure}

Our proposal for the holographic multi-classical correlation is summarized in Fig.~\ref{fig:multi-J}. Namely, we substitute the entropy part $S(A)$ with the multi-entropy $S(A:B;C)\equiv S(A:B:CD)$ as if $\rho_{ABC}$ is a pure state. In other words, we find the multi-entropy surface homologous to both $A$ and $B$ while the boundaries of $C$ with the purifying subsystem $D$ are ignored, implying the multi-entropy surface is not homologous to $C$ only. Thus, this mixed-state multi-entropy is not symmetric under the exchange $A,B\leftrightarrow C$ but only symmetric between $A\leftrightarrow B$. This is a desired feature, which is common with the entanglement entropy $S(A)\neq S(B)$ for a mixed bipartite state $\rho_{AB}$.
To emphasize the asymmetry, we denote this mixed-state multi-entropy of $\rho_{ABC}$ by $S(A:B;C)$, where the subsystems that are homologous to the multi-entropy surface are placed before the semicolon.

Alternatively, the mixed-state multi-entropy $S(A:B;C)$ can be regarded as a standard multi-entropy but $C$ and $D$ are grouped --- $S(A:B:CD)$. But we emphasize that the quantity does not depend on the purifying subsystem $D$.

The second term in the classical correlation can be considered similarly. Namely, after measuring $C$, the multi-entropy surface can anchor on the boundary of the entanglement wedge of $C$, which leads to the multipartite EWCS for $ABD$.

To summarize, our proposal for the \emph{holographic multi-classical correlation} is given by
\begin{align}
    &\phantom{==}J(A_1:\cdots:\!A_{n-1}|A_n)\nonumber\\
    &\equiv S(A_1:\cdots:\! A_{n-1}; A_n) - S(A_1:\cdots:\! A_{n-1}:E)_{A_n} \nonumber\\
    &=S(A_1:\cdots:\! A_{n-1}: A_n E) - E_W(A_1:\cdots:A_{n-1}:E),
    \label{eq:mult-J-def}
\end{align}
where the state on $A_1\cdots A_n E$ is pure and the subscript $A_n$ means the quantity is evaluated after the disentangling measurement along $\gamma_{A_n}$.

When $n=2$, it becomes $J(A_1|A_2)=S(A_1;A_2)-E_W(A_1:E)$. Recalling $S(A_1;A_2)$ is computed by the area of the minimal surface homologous to $A_1$, this is nothing but $S(A_1)$. Thus, this generalization reduces to the holographic classical correlation when $n=2$.

If all the subsystems appearing in the multi-classical correlation constitute a pure state (i.e. $E=\emptyset$), it becomes
\begin{equation}
    J(A_1\!:\!\cdots\!:\!A_{n-1}|A_n)=S(A_1\!:\!\cdots\!:\! A_n) - E_W(A_1\!:\!\cdots\!:\!A_{n-1}).
\end{equation}
The first term is given by the conventional (symmetric) multi-entropy and the second term is the multipartite EWCS for the $(n-1)$ parties.

To find the multipartite generalization of the quantum discord, we first define the multi-mutual information by changing the classical conditioning by the quantum one:
\begin{align}
    &\phantom{\rightarrow} S(A_1:\cdots:\! A_{n-1}:E)_{A_n} \nonumber\\
    &\rightarrow \frac{1}{n-1} \sum_{k=1}^{n-1} \qty(S(A_1:\cdots:\! A_k A_n:\cdots:\! A_{n-1}:E) - S(A_n)) \nonumber\\
    &\phantom{\rightarrow} \quad =: S(A_1:\cdots:\! A_{n-1}|A_n).
\end{align}
Since there are $n-1$ possibilities to consider joint multi-entropy involving $A_n$, we consider all possible multi-conditional entropies and average them.
While a different generalization of quantum conditional multi-entropy is possible, we will shortly see that with the above choice, the multi-correlation measures reduce to the bipartite one when $n=2$ and lead to the UV finiteness.

Now we can define the \emph{multi-mutual information} as
\begin{equation}
\begin{split}
    &\phantom{==}I(A_1:\cdots:\!A_{n-1}|A_n) \\
    &\equiv S(A_n) - S(A_1:\cdots:A_{n-1}|A_n)\\
    &= S(A_n)+ S(A_1:\cdots\! A_{n-1}; A_n) \\
    & \phantom{==}\, - \frac{1}{n-1} \sum_{k=1}^{n-1} S(A_1:\cdots:\! A_k A_n:\cdots:\! A_{n-1}:E).
\end{split}
\label{eq:hol-multi-I}
\end{equation}
Here we use $|$ to separate $A_n$ from the rest to show this quantity is not invariant under permutation involving $A_n$.

The definition of the multi-mutual information leads us to define the \emph{holographic multi-quantum discord} a
\begin{align}
    &\phantom{==}D(A_1:\cdots:\! A_{n-1}|A_n) \nonumber\\
    &\equiv  I(A_1:\cdots:\! A_{n-1}|A_n)-J(A_1:\cdots:\! A_{n-1}|A_n) \nonumber\\
    &= S(A_n) - \frac{1}{n-1} \sum_{k=1}^{n-1} S(A_1:\cdots:\! A_k A_n:\cdots:\! A_{n-1}:E) \nonumber\\
    &\qquad + E_W(A_1:\cdots :\! A_{n-1}:E).
    \label{eq:multi-QD-def}
\end{align}

Recalling that the multi-entropy reduces to the entanglement entropy in the pure bipartite limit, the holographic multi-mutual information, multi-classical correlation, and multi-quantum discord reduce to the bipartite ones when $n=2$.

\subsection{UV finiteness of holographic multi-correlations}
The above choice of quantum conditioning leads to UV-finite correlation measures (unless the subsystems are contiguous). This can be confirmed by checking the number of the boundary points each surface anchors. 

~

\paragraph{Holographic multi-classical correlation.}
Let us first examine UV divergence of $J(A_1:\cdots:A_{n-1}|A_n)$, whose definition is given in~\eqref{eq:mult-J-def}. The number of UV-divergent contributions in the first term, $S(A_1:\cdots: A_{n-1}; A_n)=S(A_1:\cdots: A_{n-1}:A_n E)$, is
\begin{align}
    &\phantom{=}\abs{\partial A_1\cup\cdots\cup\partial A_{n-1} \cup \partial(A_n E)} \nonumber\\
    &= \abs{[\cup_{i=1}^{n-1} \partial A_i] \cup \partial A_n \cup\partial E \setminus [\partial A_n \cap \partial E]} \nonumber\\
    &= \abs{[\cup_{i=1}^{n-1} \partial A_i] \cup \partial A_n \cup\partial E} - \abs{\partial A_n \cap \partial E}.
    \label{eq:multi-J-div1}
\end{align}
where we denote the (codimension-three) area of an entangling surface on the boundary by $\abs{\cdot}$.
Note that $\cup$ does not take multiplicity into account. For example, if $A$ and $B$ are two adjacent intervals sharing one point in AdS$_3$/CFT$_2$, the number of the boundary points $\abs{\partial A\cup \partial B}$ equals $3$, not $\abs{\partial A}+\abs{\partial B}=4$.

The number of UV-divergent contributions in the second term, $E_W(A_1:\cdots:A_{n-1}:E)=S(A_1:\cdots:\! A_{n-1}:E)_{A_n}$, is
\begin{align}
    \abs{[\cup_{i=1}^{n-1} \partial A_i] \cup\partial E} - \abs{\partial A_n}.
    \label{eq:multi-J-div2}
\end{align}
Note that the subtraction occurs because the minimal web can attach to the edge of the entanglement wedge and there should be no UV divergence from $\partial A_n$, the boundary of the measured subsystem.

Recall that $\{A_1,\cdots,A_n,E\}$ cover the entire boundary. This implies
\[
\abs{[\cup_{i=1}^{n-1} \partial A_i]\cup \partial A_n \cup\partial E} =\abs{[\cup_{i=1}^{n-1} \partial A_i] \cup\partial E}.
\]
Now, subtracting \eqref{eq:multi-J-div2} from \eqref{eq:multi-J-div1}, we find that the number of UV-divergent terms in $J(A_1:\cdots:A_{n-1}|A_n)$ is given by
\begin{equation}
    \abs{\partial A_n} - \abs{\partial A_n \cap \partial E},
    \label{eq:div-final-J}
\end{equation}
which is a portion of the boundary area of the measured subsystem $A_n$ that is not congruent to a purifier $E$.
Thus, as long as $A_n$ (the measured subsystem) is separated from the rest of the subsystems of interest ($A_1,\cdots,A_{n-1}$) by $E$, the holographic multi-classical correlation is UV-finite.\footnote{When the measured subsystems are adjacent to the other subsystems (except the purifier), $J(A_1:\cdots:A_{n-1}|A_n)$ becomes UV divergent. This is same as the bipartite mutual information, where two adjacent subsystems have a UV-divergent mutual information in quantum field theories due to an unlimited amount of entanglement across the touching boundaries. We will also see the same divergence in the holographic multi-mutual information and holographic multi-quantum discord.}

\paragraph{Holographic multi-mutual information.}
Next, let us consider holographic multi-mutual information, whose definition is given in \eqref{eq:hol-multi-I}. The number of UV divergent contributions there is
\begin{align}
    &\phantom{=}\abs{\partial A_n} + \abs{[\cup_{i=1}^{n-1} \partial A_i]\cup\partial (A_n E)} \nonumber\\
    &\qquad\qquad - \frac{1}{n-1}\sum_{k=1}^{n-1}\abs{[\cup_{i=1,i\neq k}^{n-1}\partial A_i]\cup \partial(A_k A_n) \cup \partial E}\nonumber\\
    &= \abs{\partial A_n} + \abs{[\cup_{i=1}^{n} \partial A_i]\cup\partial E} - \abs{\partial A_n\cap \partial E} \nonumber\\
    &\ - \frac{1}{n-1}\qty[(n-1)\abs{\qty(\bigcup_{i=1}^{n} \partial A_i)\cup\partial E}-\qty(\sum_{k=1}^{n-1}\abs{\partial A_k \cap \partial A_n})] \nonumber\\
    &=\frac{n}{n-1}(\abs{\partial A_n}-\abs{\partial A_n \cap \partial E}).
    \label{eq:multi-I-div}
\end{align}
In the last line, we used $\abs{\partial A_n}=\sum_{k=1}^{n-1}\abs{\partial A_k \cap \partial A_n}+\abs{\partial E \cap \partial A_n}$.

From \eqref{eq:multi-I-div}, we find that up to the overall coefficient $\frac{n}{n-1}$, the UV divergence structure is the same as the $J(A_1:\cdots:\!A_{n-1}|A_n)$. Thus, the holographic multi-mutual information is UV-finite as long as a purifier $E$ isolates the measured subsystem $A_n$ apart from the others.

\paragraph{Holographic multi-quantum discord.}
The number of UV divergence in the holographic multi-quantum discord~\eqref{eq:multi-QD-def} can be calculated by subtracting \eqref{eq:div-final-J} from \eqref{eq:multi-I-div}:
\[
\frac{1}{n-1}(\abs{\partial A_n}-\abs{\partial A_n \cap \partial E}).
\]
Thus, the holographic multi-quantum discord is also UV-finite as long as the purifier $E$ separates $A_n$ from the rest.

\section{Conclusion}\label{sec:concl}
\subsection{Summary}
To explore quantum correlations beyond entanglement, we constructed holographic duals of classical correlation and quantum discord. Table~\ref{tab:corr} shows a summary of bipartite correlation measures examined here. Based on the earlier proposal of the holographic locally accessible information based on the geometric optimization, we proposed the bulk duals for the classical correlation and quantum discord, denoted by $J_W$ and $D_W$, respectively. Since the dual boundary theory is expected to be strongly-coupled and infinite-dimensional, this is a huge departure from the existing studies, where the quantum discord has been only studied in small systems or Gaussian cases. The bulk formulae are given by the linear combination of the holographic entanglement entropy and the EWCS, circumventing any complications due to an optimization in the definitions. We analyzed their properties in detail and found all the properties of the original correlation measures are satisfied by the bulk duals, supporting our conjectures. Some of the properties lead to novel bounds for the EWCS or the Markov gap by the conditional quantities.

\begin{table*}[t]
  \centering
  \begin{tabular}{@{}lll@{\hspace{-20pt}}l@{}}
    \toprule
    \textbf{Quantity} & \textbf{Definition} & \textbf{Bulk dual} & \textbf{Other boundary duals} \\
    \midrule
    $J(A|B)$
      & Classical correlation (CC)
      & $J_{W}(A|B)$ 
      & 1WAY one-shot distillable entanglement $E_D$, \\
      & & & reflected classical correlation $J_{R}$ \\
    $D(A|B)$
      & Quantum discord, $I-J$
      & $D_{W}(A|B)$ 
      & Reflected discord~$D_{R}$ \\
    $E_{sq}(A:B)$
      & Squashed entanglement
      & $I(A:B)/2$
      & — \\
    $h$
      & Markov gap, $S_{R}-I$
      & $2E_{W}-I$
      & — \\
    $\Delta Q(A|B)$
      & NEQC, $D - E_{sq}$
      & $\Delta Q_{W}(A|B)$ 
      & Markov gap~$h(A:C)$, reflected NEQC $\Delta Q_{R}$,\\
      & & & 1WAY one-shot undistillable entanglement \\
    $I^{TC},J^{TC},D^{TC}$
      & Total correlation (TC)-based
      & —
      & — \\
    $I^{DTC},J^{DTC},D^{DTC}$
      & Dual total correlation (DTC)-based
      & —
      & — \\
    $J^{seq},D^{seq}$
      & Sequential measurements-based
      & —
      & — \\
    $I_n,J_n,D_n$
      & $n$-partite information-based
      & —
      & — \\
    $S(A:B:C)$
      & Multi-entropy for pure $\rho_{ABC}$
      & tri-way cut \\
      & & \phantom{a}among $A,B,C$
      & — \\
    $S(A:B;C)$
      & Multi-entropy for mixed $\rho_{ABC}$, 
      & tri-way cut \\
      & \phantom{a}$S(A:B:CE)$ & \phantom{a}among $A,B,CE$
      & — \\
    $E_W(A:B:C)$
      & Multipartite EWCS
      & tri-way cut within \\
      & & \phantom{a}the entanglement wedge\\
      & — \\
    $S(A:B|C)$
      & Multi-conditional entropy,\\
      & \phantom{a}$\displaystyle\frac{S(AC:B:E)+S(A:BC:E)}{2}-S(C)$\\
      &\\
    $I(A:B|C)$
      & Multi-mutual information,\\
      & \phantom{a}$S(C)+S(A:B;C)$\\
      & \phantom{a}$-\displaystyle\frac{S(AC:B:E)+S(A:BC:E)}{2}$
      & — \\
      & — \\
    $J(A:B|C)$
      & Multi-CC
      & $S(A\!:\!B;C)-E_W(A\!:\!B\!:\!E)$\\
      & — \\
    $D(A:B|C)$
      & Multi-quantum discord
      & $I(A\!:\!B|C)-J(A\!:\!B|C)$\\
      & — \\
    \bottomrule
  \end{tabular}
  \caption{List of correlation measures and their bulk/boundary duals. $E$ denotes a purifying partner. For the multi-entropic measures, tripartite cases are shown as examples.}
  \label{tab:corr}
\end{table*}

To further examine their properties, we computed $J_W,D_W$ in three-dimensional asymptotically AdS spacetimes, including Poincar\'e AdS, a two-sided eternal black hole, and a one-sided typical black hole. We identified five different phases as we increased the subsystem size, showing the discrepancy among $J_W,D_W$ and holographic squashed entanglement $E_{sq}$ in the intermediate regimes. The overall behavior was reproduced in Haar random states while the intermediate phases were interpreted as a consequence of an internal bulk structure. 
Another example examined here is a one-sided black hole. In the boundary theory, it is dual to the Gibbs state. As we increase the temperature, it
decoheres to a maximally mixed state. Naively, one expects a monotonically decreasing $D_W$ as the temperature increases since the quantum discord originates from quantum coherence. However, we provided a counterexample at the intermediate temperature above the Hawking-Page threshold. This aligns with the observation in the two-qubit Heisenberg model and is similar to the anti-Unruh effect reported in earlier literature. Our result suggests that this decoherence-against increase of discord is a universal phenomenon regardless of the dimensions or interactions of the theory. Our result also implies that the sudden death of entanglement should be restated as that of distillable entanglement.

To quantify quantum correlations beyond entanglement, we defined non-entanglement quantum correlation (NEQC) by $\Delta Q=D-E_{sq}$. Generally, this quantity is not computable as both terms involve optimizations whose closed form is not known. We overcame this by considering the bulk duals. We showed that both in holography and Haar random states in the large dimension limit, $\Delta Q$ becomes non-negative, indicating the quantum discord is quantitatively larger than entanglement. In addition, following the conjecture of holographic squashed entanglement, we found the holographic NEQC is related to the Markov gap, signaling its deep relation to tripartite entanglement and a novel interpretation of the Markov gap as an excess quantum correlation than entanglement. Viewing it from a different angle, we also showed that the NEQC equals the undistillable bipartite entanglement in holography, whose positivity also indicates the existence of the non-GHZ tripartite entanglement.

Motivated by the holographic formulae, we also looked for better correlation measures that are computable in boundary quantum systems without optimization. We defined new correlation measures $J_R,D_R,\Delta Q_R$ based on the reflected entropy and they are shown to satisfy analogous properties as the original ones. We demonstrated its validity with a family of two-qubit states and observed the reflected correlation measures capture the functional form of the original measures compared to the other existing measures.

Finally, we discussed five different generalizations of the bipartite classical correlation and quantum discord to a multipartite setting. In particular, their holographic generalizations are attractive as 1) they reduce to the original bipartite measures, 2) they are UV-finite, and 3) they are based on the multi-entropy, which is naturally obtained by generalizing the holographic entanglement entropy formula to multiple parties. While considering the holographic multi-correlation measures, we found a natural extension of the multi-entropy for mixed states, respecting the homology condition of the entropy. It is an interesting direction to compute this holographic generalization in generic quantum systems by utilizing the boundary duals of the multi-entropy and compare to the existing multipartite entanglement measures.


\subsection{Outlook}
Let us conclude this paper with possible applications in quantum gravity and quantum information. These include multipartite generalizations of correlation measures, reflected versions, and characterizing observers in quantum gravity.

~

\paragraph{Multipartite entanglement and correlations.}--- The nonzero NEQC or undistillable entanglement signals the existence of the tripartite entanglement in holography and Haar random states as suggested previously~\cite{Mori:2024gwe,Li:2025nxv,Akers:2019gcv,Iizuka:2025bcc}. It is interesting if this connection can be made more quantitative, namely, whether the NEQC bounds a tripartite entanglement measure. Furthermore, to identify the genuine measure for the multipartite correlations, a further comparison and analysis of their properties need to be carried out. As shown for the bipartite cases, it is tempting to think a multipartite quantum discord upper bounds faithful entanglement measures. 
While the definition of the faithful multipartite entanglement measures is still an ongoing debate~\cite{Basak:2024uwc}, one possible definition is the multipartite squashed entanglement~\cite{Yang_2009,Avis_2008}, whose holographic dual may be merely given by the total correlation~\cite{Umemoto:2018jpc}.
Another possibility is to define the genuine multipartite entanglement by removing $(n-1)$-partite correlations from the $n$-partite correlations as done in~\cite{Iizuka:2025ioc,Iizuka:2025caq}. Since these previous works are based on holographic multi-entropies, they can be applied to our holographic generalization of the multipartite correlation measures straightforwardly. 

Another interesting avenue is the relation to the distillable entanglement and multipartite entanglement. Observing that the bipartite classical correlation is identified with the (one-way) distillable entanglement, it is worth thinking if the multipartite classical correlation is identified with the multipartite distillable entanglement. As of now, even the definition of distillable multipartite entanglement is not clear; thus, analyzing the multipartite classical correlations and distillation protocols based on them may be useful to give a well-defined notion of multipartite entanglement and its distillation.

~

\paragraph{Multipartite Markov gap.}--- It is worth noting that our approach may give a natural generalization of the Markov gap, overcoming the limitation mentioned in~\cite{Iizuka:2025ioc}. The multipartite Markov gap may be defined as
\begin{equation}
    h(A_1:\cdots:A_{n-1}:E) \equiv 2\Delta Q(A_1:\cdots:A_{n-1}|A_n)
\end{equation}
where the multipartite NEQC is defined as $D(A_1:\cdots:A_{n-1}|A_n)$ minus the multipartite squashed entanglement.
Since the bipartite Markov gap can distinguish the GHZ entanglement from others, it is interesting to know if this multipartite Markov gap can distinguish a certain class of multipartite entanglement.

~

\paragraph{Observer in quantum gravity.}--- It is believed that the algebra of quantum gravity is free from UV divergences. Effective field theories of a bulk local region, such as one side of a two-sided black hole or static patch of de Sitter spacetime, naively suffer from the UV divergence. In other words, its algebra is type III von Neumann algebra. By introducing observer(s), it is thought to render the algebra to type II (or possibly type I)~\cite{Witten:2021unn,Chandrasekaran:2022cip}. Typically, an observer is treated semiclassically as a classical-quantum state $\rho_{AB}$~\cite{Witten:2021unn,Jensen:2023yxy,DeVuyst:2024uvd,Harlow:2025pvj}. This is natural from the viewpoint that an observer carries a clock, which forms an approximately orthonormal basis. Notably, the quantum discord $D(B|A)$ becomes zero for classical-quantum states. Thus, the quantum discord could identify the observers' degrees of freedom in the bulk effective field theory. Since we know the holographic formula, the effective field theory can be even strongly coupled. Moreover, since the holographic quantum discord is written as a relative entropy, it could be defined via the relative modular operator in any von Neumann algebras even when the notion of entropy or reduced density matrices is absent.

Another point of view regarding an observer is that it is a detector performing a measurement to record its `time' in its clock state. Such a case where measurements or more generally LOCC is performed in a gravitational system through an external system is often considered in quantum reference frames~\cite{Page:1983uc,PhysRevD.30.368,Giovannetti:2015qha,Lake:2023nua} and black hole interior reconstruction through islands~\cite{Qi:2021sxb}. Studying holographic discord in these examples may sharpen the observer's role based on the correlation structure.

~

\paragraph{Reflected correlation measures.}--- While the existence of quantum correlation beyond entanglement is illuminating, its nature has been mysterious over decades. Many proposed measures have difficulties either in their interpretations and/or computational feasibility. For example, the geometric discord can be calculated easily~\cite{Banerjee:2023liw} but it can exceed entanglement~\cite{Rana_2012} and increase under local operations~\cite{PhysRevA.86.034101,Torun:2024djb}.
The definition of the quantum discord involves an optimization over all possible POVM. This makes its computation in larger systems impossible. To the authors' knowledge, when the rank becomes more than two, there is no known formula. If one restricts to the Gaussian states and Gaussian POVM, one can obtain a closed expression, however, it is not possible for strongly-coupled systems. Our proposal of the holographic formula is advantageous in these two aspects. The boundary theory is a field theory so it is infinite-dimensional in the continuum limit and strongly-coupled. However, one limitation is the quantum system needs to be holographic. Instead, one could consider using the reflected correlation measures. Although they may depart from the true correlation measures, they are expected to behave similarly and overwhelm other existing measures, as demonstrated with two-qubit states in this paper. It is interesting to study how they behave in quantum many-body systems to probe their nature beyond entanglement.

 \begin{acknowledgments}
 \vspace{2em}
 \noindent The author would like to thank Rathindra Nath Das, Masahiro Hotta, Ling-Yan Hung, Kyan Louisia, Donald Marolf, Masamichi Miyaji, Shintaro Minagawa, Masanao Ozawa, Shang-Ming Ruan, Toshihiko Sasaki, Tadashi Takayanagi, Kotaro Tamaoka, Izumi Tsutsui, Toyohiro Tsurumaru, Tomonori Ugajin, Herbie Warner, Zixia Wei, Beni Yoshida, and Yijian Zou
 for useful discussions and comments. This research was supported in part by the Perimeter Institute for Theoretical Physics, SOKENDAI, and the Atsumi Scholarship from the Atsumi International Foundation. Research at Perimeter Institute is supported by the Government of Canada through the Department of Innovation, Science and Economic Development and by the Province of Ontario through the Ministry of Research, Innovation and Science. This work was supported by JSPS KAKENHI Grant Number 23KJ1154, 24K17047.
 \end{acknowledgments}

\appendix

\section{Calculation of holographic correlation measures for TFD states}\label{app:BTZ}
Let us calculate the classical and quantum correlations for the TFD state~\eqref{eq:TFD}. The boundary has topology $S^1\times\mathbb{R}$ and each subsystem is specified by its angle.
Given a subregion $A$ of angle $l$, its holographic entropy is given by
\begin{equation}
	S_A =
	\begin{cases}
		S[l] &\quad (l\le 2\pi - l_\ast)\\
		S_{BH}+S[2\pi-l] &\quad (l\ge 2\pi - l_\ast)
	\end{cases}
,
\end{equation}
where
\begin{equation}
    l_\ast = \pi -\frac{\beta}{2\pi} \log \cosh\frac{2\pi^2}{\beta}<\pi
    \label{eq:l-ast}
\end{equation}
and
\begin{equation}
	S[l] = \frac{c}{3} \log \qty(\frac{\beta}{\pi \epsilon} \sinh\frac{l \pi}{\beta}),\quad S_{BH}= \frac{c}{3}\frac{2\pi^2}{\beta}.
\end{equation}
For the subsystem $B$, $S_B=S_{BH}$. For the subsystem $C$,
\begin{equation}
	S_C = 
	\begin{cases}
		S_{BH}+S[l] &\quad (l\le l_\ast)\\
		S[2\pi-l] &\quad (l\ge l_\ast)
	\end{cases}
.
\end{equation}
Subsequently, the mutual information is given by
\begin{equation}
	I(A:B) = 
	\begin{cases}
		0 \quad & (l\le l_\ast) \\
		S_{BH} + S[l] - S[2\pi-l] \quad & (l_\ast \le l \le 2\pi-l_\ast)\\
		2S_{BH} \quad & (l \ge 2\pi-l_\ast)
	\end{cases}
	.
\end{equation}

The EWCS between $A$ and the purification partner $C$ is given as
\begin{equation}
	E_W(A:C) = 
	\begin{cases}
		S[l] &\quad \qty(l\le l_0) \\
		S_{\mathrm{atyp}} & \quad \qty(l_0 \le l \le 2\pi - l_0) \\
		S[2\pi -l] &\quad \qty(l\ge 2\pi - l_0) 
	\end{cases}
	,
\end{equation}
where 
\begin{equation}
	S_{\mathrm{atyp}} = \frac{c}{3}\log\frac{\beta}{\pi\epsilon},\quad l_0 = \min\qty(\pi, \frac{\beta}{\pi}\log(1+\sqrt{2}))
    \label{eq:l-zero}
\end{equation}
is the entanglement entropy unique to an atypical, disentangled black hole. It is given by the area of the minimal surfaces that anchor on the horizon.
Note that regardless of the temperature, $l_\ast\le l_0$ so the transition of $E_W(A:C)$ happens after the transition of $I(A:B)$.

\begin{figure}
	\centering
	\vspace{10pt}
	\includegraphics[width=0.9\linewidth]{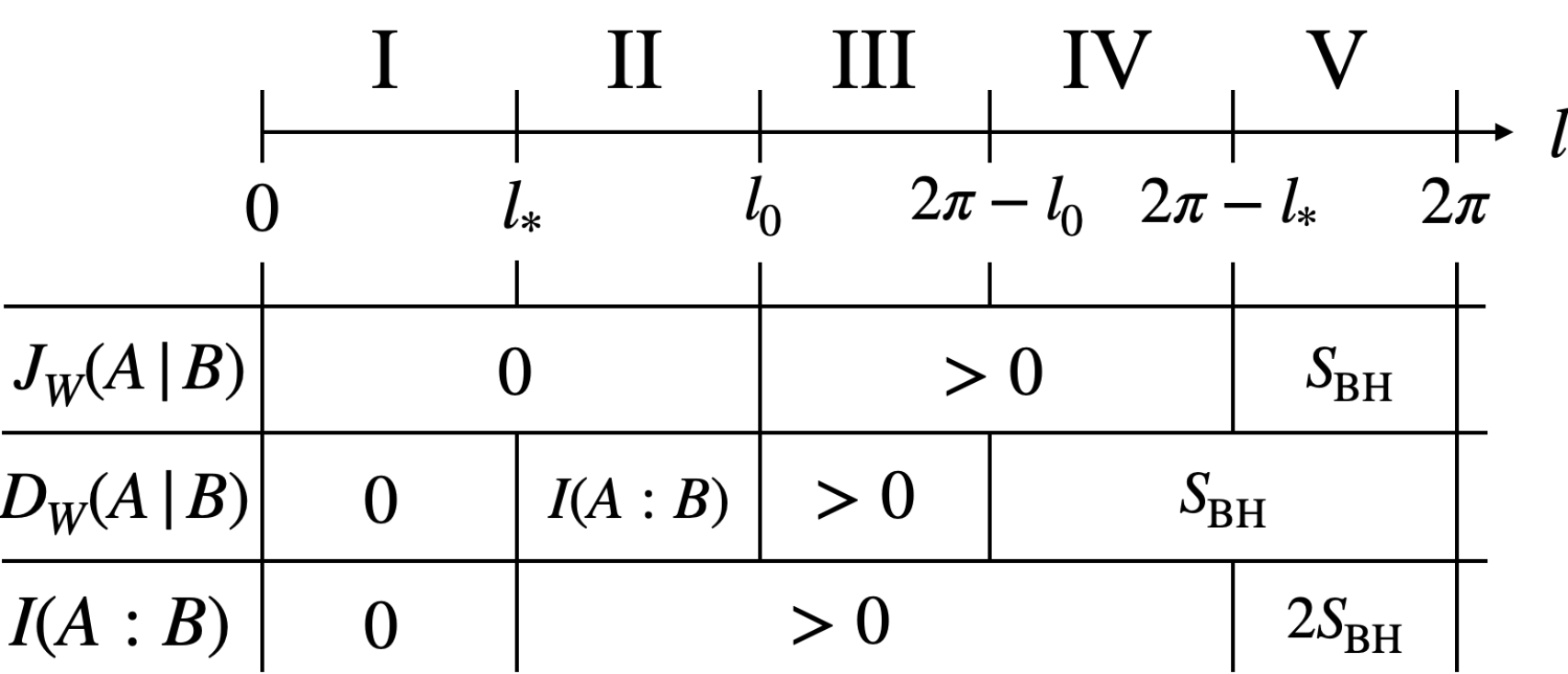}
	\caption{Depending on the value of $J_W(A|B)$ and $D_W(A|B)$, there are five distinct phases I-V.}
	\label{fig:phases}
\end{figure}

Finally, the gravity dual of the classical correlation is given by
\begin{equation}
	J_W(A|B)= \frac{c}{3}
	\begin{cases}
		0 & \ (\mathrm{I,II}) \\
		\log\qty(\sinh\displaystyle\frac{l\pi}{\beta}) & \ (\mathrm{III}) \\
		\log\qty(\displaystyle\frac{\sinh \frac{l \pi}{\beta}}{\sinh \frac{(2\pi- l)\pi}{\beta}})  & \ \qty(\mathrm{IV}) \\
		\displaystyle\frac{2\pi^2}{\beta} & \ \qty(\mathrm{V})
	\end{cases}
\end{equation}
and the gravity dual of the quantum discord is given by
\begin{equation}
	D_W(A|B) = \frac{c}{3}
	\begin{cases}
		0 & \ (\mathrm{I}) \\
		\displaystyle\frac{2\pi^2}{\beta} - \log\qty(\displaystyle\frac{\sinh \frac{(2\pi- l)\pi}{\beta}}{\sinh \frac{l \pi}{\beta}})  & \ (\mathrm{II}) \\
		\displaystyle\frac{2\pi^2}{\beta} - \log\qty(\sinh \frac{(2\pi- l)\pi}{\beta})  & \ \qty(\mathrm{III}) \\
		\displaystyle\frac{2\pi^2}{\beta} & \ \qty(\mathrm{IV,V})
	\end{cases}
.
\end{equation}
There are five distinct phases, denoted as I-V. The range of each phase and the corresponding behavior of $J_W(A|B),D_W(A|B)$ are shown in Fig.~\ref{fig:phases}.

\section{Calculation of holographic correlation measures in finite temperature}\label{app:hol-BH}
Let us consider holographic correlation measures in finite temperature $\beta^{-1}>\frac{1}{2\pi}$. The gravity dual of the Gibbs state is given by the BTZ black hole. As shown in Fig.~\ref{fig:BH-decoh}, we take two subregions $A$ and $B$ symmetrically on the boundary of the one-sided black hole.

\begin{figure}
    \centering
    \includegraphics[width=0.8\linewidth]{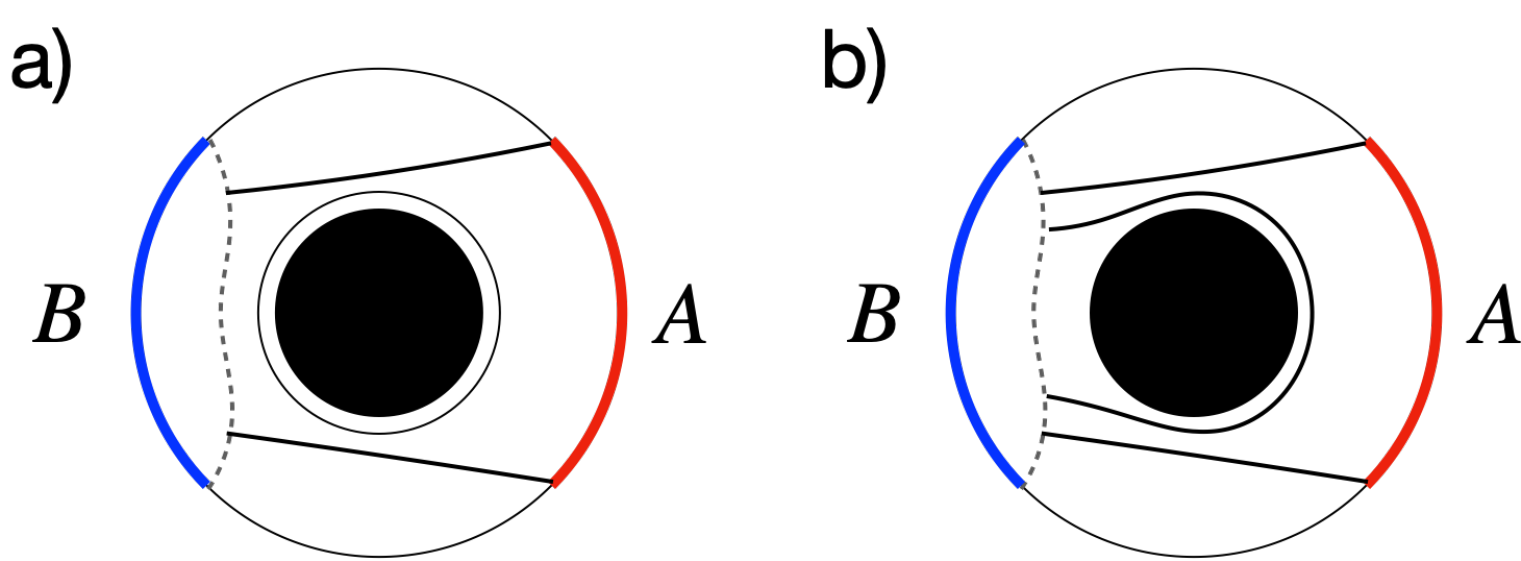}
    \caption{Nontrivial phases of $E_W(A:C)$ a) when the EWCS separating the environment is given by the black hole horizon and b) when it anchors on the extremal surface of $B$.}
    \label{fig:ewcs-phases}
\end{figure}

Among the contributions to $E_W(A:C)$, the one from a geodesic wrapping around the black hole becomes UV finite. Thus, it is important to correctly identify the UV cutoffs to reproduce the expected behavior of the classical and quantum correlations in accord with the entanglement entropy. We will follow the expression of the EWCS can be found in~\cite{Takayanagi:2017knl}. Then, to correctly reproduce the entanglement entropy, we claim the UV cutoff for $A$ and $B$ is given by $2\epsilon$ while that of the complementary subsystem is given by $\epsilon$. See~\cite{Kusuki:2022ozk} for the relevant discussion.
With this convention of the UV cutoff, the entanglement entropy of the subsystem $A$ or $B$, whose angle is $\theta$, is given by
\begin{equation}
    S_A=S_B=\frac{c}{3}\log\qty(\frac{\beta}{2\pi\epsilon}\sinh\frac{\pi \theta}{\beta}),
\end{equation}
while the entanglement entropy of the complementary region is given by
\begin{equation}
    S_{AB}=\min\qty[S_A+S_B,\frac{2c}{3}\log\qty(\frac{\beta}{\pi\epsilon}\sinh\frac{\pi(\pi-\theta)}{\beta})+S_{BH}].
\end{equation}
For the EWCS $E_W(A:C)$, there are two possible phases other than the trivial case where $E_W(A:C)=S_A$. See a) and b) of Fig.~\ref{fig:ewcs-phases} for the illustration. $E_W(A:C)$ can be computed accordingly by following~\cite{Takayanagi:2017knl}.

Now, we can calculate the holographic correlation measures for the reflection symmetrically placed subregions $A,B$ with their respective angle $\theta$. The classical correlation is given by
\begin{equation}
	J_W(A|B)=\max\qty[0,\frac{c}{3}\qty(\log \frac{\sinh^2\frac{\pi \theta}{\beta}}{4\sinh\frac{\pi(\pi-\theta)}{\beta}\sinh\frac{\pi^2}{\beta}}-s(\beta,\theta))],
\end{equation}
where
\begin{align}
    s(\beta,\theta)=\min\qty(\frac{2\pi^2}{\beta},\mathrm{arctanh}\sqrt{z}),\quad z = 1-\frac{\sinh^2\frac{\pi\theta}{\beta}}{\sinh^2\frac{2\pi^2}{\beta}}.
\end{align}
The squashed entanglement $E_{sq}(A:B)=I(A:B)/2$ is given by
\begin{equation}
	E_{sq}(A:B)=\max\qty[0,\frac{c}{3}\qty(\log\frac{\sinh\frac{\pi \theta}{\beta}}{2\sinh\frac{\pi(\pi-\theta)}{\beta}}-\frac{\pi^2}{\beta})].
\end{equation}
When $J_W(A|B)=0$, $D_W(A|B)=I(A:B)$. When $I(A:B)=0$, $D_W(A|B)=0$. Otherwise, we have $0<J_W(A|B)<I(A:B)$. In this regime, the quantum discord is given by
\begin{equation}
	D_W(A|B)= \frac{c}{3}\log\frac{\sinh\frac{\pi^2}{\beta}}{\sinh\frac{\pi(\pi-\theta)}{\beta}}+s(\beta,\theta).
\end{equation}

\section{Calculation of correlation measures for the two-qubit rank-two state}\label{app:rank2}
Let us compute $J(A|B),D(A|B)=\Delta Q(A|B)$ of $\rho_{AB}$ given in \eqref{eq:param-state}. The reduced density matrices are given as
\begin{equation}
	\rho_A= \frac{\bm{1}}{2},\quad \rho_B=\frac{1}{2}
	\mqty(1+\cos^2\theta & \sin\theta\cos\theta\\ \sin\theta \cos\theta & \sin^2\theta)
,
\end{equation} 
where the basis of the matrix representation is $\{\ket{0},\ket{1}\}$. This leads to $S_A=\log 2$ and $S_B=H(\sin^2\frac{\theta}{2})$, where $H(p)=-p\log p - (1-p)\log(1-p)$ is the binary entropy. Since $S_{AB}=\log 2$, the mutual information is given by $I(A:B)=H(\sin^2\frac{\theta}{2})$.

In general, finding the optimal measurement is hard as it could be a POVM rather than an (orthogonal) projective measurement. However, it is known that for a rank-two state, optimizing over orthogonal measurements is sufficient~\cite{Galve_2011}. This enables us to compute $J(A|B),D(A|B)$ explicitly.

Given generic orthogonal measurement operators $\{\dyad{\varphi}_B,\dyad{\bar{\varphi}}_B\}$ parametrized by $\varphi$, where
\begin{align*}
	\ket{\varphi} &= \cos\varphi \ket{0}+\sin\varphi \ket{1},\\
	\ket{\bar{\varphi}}&=\sin\varphi \ket{0} - \cos\varphi \ket{1},
\end{align*}
we can compute the post-measurement state on $A$, $\{\rho_{A|\varphi}, \rho_{A|\bar{\varphi}}\}$ and the associated probability distribution, $\{p_{\varphi}, p_{\bar{\varphi}}\}$. By maximizing $S_A- p_{\varphi} S(\rho_{A|\varphi})-p_{\bar{\varphi}} S(\rho_{A|\bar{\varphi}})$ over $\varphi$, we find $\varphi=\frac{\theta}{2}\pm\frac{\pi}{4}$ is optimal. This leads
\begin{align}
	J(A|B)&=\log 2 -H\qty(\sin^2\qty(\frac{\theta}{2}-\frac{\pi}{4})), \\
	D(A|B)&=H\qty(\sin^2\frac{\theta}{2})+H\qty(\sin^2\qty(\frac{\theta}{2}-\frac{\pi}{4})) -\log 2,
\end{align}
and $\Delta Q(A|B)=D(A|B)$ because $E_{sq}(A:B)=0$.

Alternatively, the above results can be obtained from the entanglement of formation $E_F$ via the Koashi-Winter relation~\cite{Shi_2011}. The analytical expression follows from the fact that $E_F$ of any rank-two states can be written in terms of the concurrence.

Next, let us compute the reflected correlation measures. The purification of $\rho_{AB}$ is
\begin{equation}
	\ket{\Psi}_{ABC}=\frac{1}{\sqrt{2}}\qty(\ket{000}+\ket{1\theta 1})_{ABC}.
\end{equation}
By tracing out $B$,
\begin{equation}
	\rho_{AC} = \frac{1}{2} \mqty(1 & \cos\theta \\ \cos\theta & 1).
\end{equation}
Its canonical purification is given by
\begin{equation}
	\ket*{\rho^{1/2}}_{ACA^\ast C^\ast} = \cos\frac{\theta}{2} \ket{I}_{AC}\ket{I}_{A^\ast C^\ast} + \sin\frac{\theta}{2}\ket{Z}_{AC}\ket{Z}_{A^\ast C^\ast},
\end{equation}
where $\ket{I}=\frac{1}{\sqrt{2}}(\ket{00}+\ket{11})$ and $\ket{Z}=\frac{1}{\sqrt{2}}(\ket{00}-\ket{11})$. 
The reduced density matrix becomes
\begin{equation}
	\rho_{AA^\ast}=\frac{\bm{1}+\sin\theta Z\otimes Z}{4},
\end{equation}
where $Z=\dyad{0}-\dyad{1}$.
The reflected entropy given by its von Neumann entropy becomes
\begin{equation}
	S_R(A:C)=\log 2 + H\qty(\sin^2\qty(\frac{\theta}{2}-\frac{\pi}{4})).
\end{equation}
From this, the reflected correlation measures are computed as follows:
\begin{align}
	J_R(A|B)&=\frac{1}{2} \qty(\log 2 -  H\qty(\sin^2\qty(\frac{\theta}{2}-\frac{\pi}{4}))), \\
	D_R(A|B) & =  H\qty(\sin^2\frac{\theta}{2}) + \frac{1}{2} H\qty(\sin^2\qty(\frac{\theta}{2}-\frac{\pi}{4})) - \frac{1}{2}\log 2,\\
	\Delta Q_R(A|B) & =  \frac{1}{2} \qty(H\qty(\sin^2\frac{\theta}{2}) +  H\qty(\sin^2\qty(\frac{\theta}{2}-\frac{\pi}{4})) - \log 2).
\end{align}

They behave similarly to the original, optimized measures. In particular,
\begin{align}
	J(A|B) & = \frac{J_R(A|B)}{2}\\
	D(A|B) & = 2 D_R(A|B) - I(A:B)\\
	\Delta Q(A|B) & = 2 \Delta Q_R(A|B).
\end{align} 

Finally, let us also compute the diagonal discord. It is defined as a quantum discord before the optimization, thus it upper bounds the (optimized) quantum discord. The measurement on $B$ is chosen to be a projective measurement whose basis is the eigenbasis of the reduced density matrix $\rho_B$. Based on the eigenbasis of $\rho_B$,
\begin{equation}
	\ket{\nu_1}=\mqty(\cos\frac{\theta}{2} \\ \sin\frac{\theta}{2}), \quad \ket{\nu_2}=\mqty(-\sin\frac{\theta}{2} \\ \cos\frac{\theta}{2}),
\end{equation}
the diagonal discord is given by
\begin{equation}
	D_{\rm diag}(A|B) = H\qty(\sin^2\frac{\theta}{2}).
\end{equation}
Note that the measurement in the eigenbasis of $B$ leads to zero information gain $J_{\rm diag}(A|B):={I(A:B)-D_{\rm diag}(A|B)}=0$, which means the diagonal discord is farthest from the optimal estimation.

\bibliography{ref}
\end{document}